\documentclass[fleqn,usenatbib]{mnras}

\usepackage{newtxtext,newtxmath}
\usepackage{deluxetable}
\usepackage{amsmath}
\usepackage[T1]{fontenc}

\DeclareRobustCommand{\VAN}[3]{#2}
\let\VANthebibliography\thebibliography
\def\thebibliography{\DeclareRobustCommand{\VAN}[3]{##3}\VANthebibliography}

\usepackage{graphicx}	% Including figure files
\usepackage{amsmath}	% Advanced maths commands
\title[IM Lup turbulence]{Evidence for Non-zero Turbulence in the Protoplanetary disc around IM Lup}

\author[Flaherty et al.]{
Kevin Flaherty$^{1}$\thanks{E-mail: kmf4@williams.edu (KMF)},
A. Meredith Hughes$^{2}$,
Jacob B. Simon$^{3}$,
Alicia Smith Reina$^{1}$,
Chunhua Qi$^{4}$,
Xue-Ning Bai$^{5,6}$,\newauthor
Sean M. Andrews$^{4}$,
David J. Wilner$^{4}$,
and \'{A}gnes K\'{o}sp\'{a}l$^{7,8,9,10}$
\\
$^{1}$Department of Astronomy and Department of Physics, Williams College, Williamstown, MA, USA\\
$^{2}$Van Vleck Observatory, Astronomy Department, Wesleyan University, 96 Foss Hill Drive, Middletown, CT, USA\\
$^{3}$Department of Physics and Astronomy, Iowa State University, Ames, IA, USA\\
$^{4}$Harvard-Smithsonian Center for Astrophysics, 60 Garden St., Cambridge, MA, USA\\
$^{5}$Institute for Advanced Study, Tsinghua University, Beijing 100084, People's Republic of China\\
$^{6}$Tsinghua Center for Astrophysics, Tsinghua University, Beijing 100084, People's Republic of China\\
$^{7}$Konkoly Observatory, HUNREN Research Centre for Astronomy and Earth Sciences,
MTA Centre of Excellence, Konkoly-Thege Mokl\'{o}s \'{u}t 15-17, 1121 Budapest, Hungary\\
$^{8}$Insitute of Physics and Astronomy, ELTE E\"otv\"os Lor\'and University, P\'azm\'any P\'eter s\'et\'any 1/A, 1117 Budapest, Hungary\\
$^{9}$Max Planck Institute for Astronomy, K\"{o}nigstuhl 17, 69117 Heidelberg, Germany\\
}

\date{Accepted XXX. Received YYY; in original form ZZZ}

\pubyear{2023}

\graphicspath{{./}{figures/}}
\begin{document}
\label{firstpage}
\pagerange{\pageref{firstpage}--\pageref{lastpage}}
\maketitle

\begin{abstract}
The amount of turbulence in protoplanetary discs around young stars is critical for determining the efficiency, timeline, and outcomes of planet formation. It is also difficult to measure. Observations are still limited, but direct measurements of the non-thermal, turbulent gas motion are possible with the Atacama Large Millimeter/submillimeter Array (ALMA). Using CO(2--1)/$^{13}$CO(2--1)/C$^{18}$O(2--1) ALMA observations of the disc around IM Lup at $\sim0\farcs4$ ($\sim$60 au) resolution we find evidence of significant turbulence, at the level of $\delta v_{\rm turb}=(0.18-0.30)$c$_s$. This result is robust against systematic uncertainties (e.g., amplitude flux calibration, midplane gas temperature, disc self-gravity). We find that gravito-turbulence as the source of the gas motion is unlikely based on the lack of an imprint on the rotation curve from a massive disc, while magneto-rotational instabilities and hydrodynamic instabilities are still possible, depending on the unknown magnetic field strength and the cooling timescale in the outer disc.
\end{abstract}


%%%%%%%%%%%%%%%%%%%%%%%%%%%%%%%%%%%%%%%%%%%%%%%%%%

%%%%%%%%%%%%%%%%% BODY OF PAPER %%%%%%%%%%%%%%%%%%

\section{Introduction}\label{sec:intro}
Turbulence is a primary factor in the planet formation process, affecting chemical evolution \citep{semenov2011,furuya2014,xu2017}, collisional growth of dust grains \citep{ormel2007,birnstiel2010}, and the vertical/radial concentration of dust grains \citep{dullemond18,rosotti2020,jennings2022}. Turbulence affects the structures generated in a disc by a newly formed giant planet \citep{bae2018}, and any uncertainty in the turbulence level influences our ability to interpret the myriad of complex structures that have been revealed among many protoplanetary discs \citep[e.g.,][]{andrews2018,long2018,oberg21}. As such, understanding the strength of turbulence, and how it varies from system to system, is important in understanding the conditions under which planets form and evolve. 

Often quantified as $\alpha$ in the context of the $\alpha$-disc model of viscosity \citep[$\nu=\alpha c_s H$ where $\nu$ is the viscosity, $c_s$ is the sound speed, and $H$ is the pressure scale height][]{shakura1973}, early studies of protoplanetary disc evolution inferred a global value of $\alpha\sim10^{-2}$ based on the measured accretion rate onto the central star \citep{hartmann1998}. More recent observations of dust vertical/radial diffusion \citep{pinte2016,dullemond18,rosotti2020,jennings2022,ohashi2019,ueda2021,doi2021,Franceschi2022,Villenave2022}, disc sizes \citep{najita18,trapman20,long2022}, and the relation between accretion rate and disc mass/radius as predicted by viscous evolution \citep{ansdell2018,ribas2020} have found $\alpha=10^{-4}-10^{-3}$, suggesting more modest levels of turbulence, at least in the outer parts of the disc\footnote{see \citet{Lesur2022} and \citet{pinte2022} for a more detailed discussion of $\alpha$, and how not all $\alpha$ measurements are the same.}. 

%%% Why did Hartmann et al. 1998 measure higher alpha values?? 

Dust diffusion and angular momentum exchange are valuable but indirect methods for measuring turbulent gas motion. The Doppler shifts of molecular line emission directly traces the kinematics of the gas. Early results \citep{guilloteau2012,hughes2011} were limited in spatial resolution and sensitivity, but ALMA has dramatically improved on this capability. Observations of the discs around HD 163296 \citep{flaherty2017}, TW Hya \citep{Teague18,flaherty18}, and V4046 Sgr and MWC 480 \citep{flaherty20} placed stringent upper limits on the amount of turbulence, constraining the non-thermal velocity component to $<$0.05c$_s$ -- $<$0.15c$_s$ between different targets ($\alpha$ less than a few $\times10^{-3}$, assuming $\alpha\sim (\delta v_{\rm turb}/c_s)^2$ and that $\delta v_{\rm turb}/c_s$ is constant throughout the disc). The exception to the general trend of weak turbulence was DM Tau, which displayed turbulence of (0.18 -- 0.28)c$_s$ \citep{flaherty20}. 

Why some systems are turbulent while others are not depends on the physical mechanism(s) that can drive turbulence within a protoplanetary disc. Gravito-turbulence \citep{shi14,forgan12} relies on gravitational instabilities within the disc, and hence different turbulence velocities could reflect different disc masses. The magneto-rotational instability \citep[MRI,][]{balbus1998} relies on a coupling between magnetic fields and partially ionized gas, and variations in the turbulent velocities may reflect either differences in the magnetic field geometry and/or strength or in the ionization of the disc gas \citep{simon18}. Hydrodynamic instabilities, e.g., the Vertical Shear Instability \citep{Nelson2013} which relies on the change in orbital velocity with height within the disc, depend on the cooling timescale \citep{lyra2021,Lesur2022}, which in turn may reflect the properties of the dust population in the outer disc.

Molecular line studies are limited in size, and often focus on the brightest discs given the high SNR required to measure the $\sim$100 m s$^{-1}$ non-thermal motions characteristic of turbulence in protoplanetary discs. This limits the number of sources in which we can distinguish between turbulence and no/weak turbulence. With this caveat in mind, DM Tau does stand out from the rest of the sources in the molecular line sample as being the youngest system, hinting at a connection between turbulence and age. A change in turbulence with age may reflect changes in disc mass \citep{manara2022}, magnetic field strength \citep{simon18,weiss21}, or cooling timescale, depending on the model for turbulence. A decrease in turbulence with age could affect the turbulence that the planets encounter as they migrate through the gas rich disc \citep{huhn2021}, with weaker turbulence at later times leading to more compact planetary systems with orbits close to resonances. 

Here we examine the young source IM Lup, around which we find evidence of non-zero turbulence. At an age of $\sim$1 Myr \citep{alcala17,mawet12} it is comparable in age to DM Tau. It also has a large CO disc; \citet{long2022} find that DM Tau (876 au) and IM Lup (803 au) have two of the three largest CO discs among the 44 targets they study, with the majority of discs having radii between 50 and 350 au. Such large discs, when combined with the young ages, are consistent with substantial viscous evolution \citep{trapman20,long2022}, which may be a sign of turbulence. More recently, \citet{paneque-carreno2023}
 examine the non-thermal motion with the CN emission line and find evidence of turbulence at (0.4--0.6)c$_s$.
%Here we examine the young source IM Lup. At an age of $\sim$1 Myr \citep{alcala17,mawet12} it is comparable in age to DM Tau and its large CO disc \citep{trapman20,long2022} may be a sign of substantial viscous evolution, consistent with large turbulence. 

In section~\ref{sec:data} we describe the data and models used to constrain turbulence, while we discuss the finding of non-zero turbulence around IM Lup in section~\ref{sec:results}. In section~\ref{sec:discussion} we tackle the question of what could be driving the turbulence around DM Tau and IM Lup, and what makes these systems different from the other sources, in order to better understand the expected range of turbulence levels among protoplanetary discs, and to constrain the physical mechanism driving the turbulence. 

%van Terwisga et al. 2018 find that 2% (IM Lup and V1094 Sco) of Lupus discs have continuum emission at radii larger than 200 au. 
%Long et al. 2022: GO Tau and DM Tau have the two largest CO discs, at 1014 and 876 au respectively. Next largest is IM Lup at 803 au, followed by DL Tau at 597 au. Most are between 50 and 350 au. 

\section{Data \& Model} \label{sec:data}
We use data from project 2013.1.00798.S (PI: C. Pinte), originally presented in \citet{pin18}. These data consist of six spectral windows, and in this work we focus on the spectral windows centered on CO(2--1), $^{13}$CO(2--1), and C$^{18}$O(2--1). These spectral windows are centered at 230.54 GHz, 220.394 GHz, and 219.556 GHz respectively, with a channel spacing of 30.518 kHz, corresponding to a velocity spacing of 39 m s$^{-1}$, 41 m s$^{-1}$, and 42 m s$^{-1}$ for the three emission lines.

The raw data were run through the \texttt{CASA} pipeline v4.7.2, while analysis was performed in \texttt{CASA} v5.4.0. Self-calibration was performed using a continuum spectral window with two rounds of phase calibration. For CO(2--1), images are generated using Briggs weighting with robust=0.5, resulting in a beam size of 0$\farcs$3$\times$0$\farcs$5 and an rms of 8.6 mJy bm$^{-1}$ channel$^{-1}$. Natural weighting was used for $^{13}$CO and C$^{18}$O resulting in a beam size of 0$\farcs$4$\times$0$\farcs$7 for both lines, with an rms of 7.9 mJy bm$^{-1}$ channel$^{-1}$ and 5.6 mJy bm$^{-1}$ channel$^{-1}$ for $^{13}$CO and C$^{18}$O respectively.
%0.7x0.4" beam size for 13CO, 0.7"x0.4" for C18O
%7.9mJy/bm noise for 13CO, 5.6 mJy/bm noise for C18O

To model the molecular line emission, we use the parametric surface density and disc temperature structure as described in \citet{flaherty20} and references therein. To briefly summarize, the surface density is assumed to follow a power law with an exponential tail:
\begin{equation}
\Sigma_{\rm gas}(r) = \frac{M_{\rm gas}(2-\gamma)}{2\pi R^2_c}\left(\frac{r}{R_c}\right)^{-\gamma}\exp\left[-\left(\frac{r}{R_c}\right)^{2-\gamma}\right],
\end{equation}
where $M_{\rm gas}$, $R_c$, and $\gamma$ are the gas mass (in M$_{\odot}$), critical radius (in au) and power law index respectively. The disc extends from $R_{\rm in}$ to 1000 au. 

The temperature structure is parameterized as a power law with radius, with a smooth function connecting the cold midplane and the warm atmosphere. %\begin{eqnarray}
%T_{\rm mid} & = & T_{\rm mid0}\left(\frac{r}{150\ \rm au}\right)^{q}\\
%T_{\rm atm} & = & T_{\rm atm0}\left(\frac{r}{150\ \rm au}\right)^{q}\\
%T_{\rm gas}(r,z) & = & \left\{
%\begin{array}{ll}
%T_{\rm atm} + (T_{\rm mid}-T_{\rm atm})(\cos\frac{\pi z}{2Z_q})^{2} & \mbox{if $z < Z_q$} \\
%T_{\rm atm} & \mbox{if $z \ge Z_q$}
%\end{array}
%\right.\\
%Z_q & = & Z_{q0} (r/150\ {\rm au})^{1.3}
%\end{eqnarray}
\begin{equation} 
\begin{aligned}
    \begin{split}
    T_{\rm gas}(r,z<Z_q) & = & \\
    & T_{\rm atm0}\left(\frac{r}{150\ \rm au}\right)^{q} \\
    & + (T_{\rm mid0}\left(\frac{r}{150\ \rm au}\right)^{q}- T_{\rm atm0}\left(\frac{r}{150\ \rm au}\right)^{q}))\cos\frac{\pi z}{2Z_q})^{2} \end{split} \\
\end{aligned}
\end{equation}
\begin{equation}
    \begin{aligned}
    T_{\rm gas}(r,z>Z_q) & = & T_{\rm atm0}\left(\frac{r}{150\ \rm au}\right)^{q} & \mbox{if $z > Z_q$} \\
    Z_q & = & Z_{q0} (r/150\ {\rm au})^{1.3}
    \end{aligned}
\end{equation}
The parameter $Z_q$ is the height above the midplane at which the gas temperature reaches its maximum value, and $Z_{q0}$ was set to twice the local pressure scale height \citep{dartois2003}, with the pressure scale height defined as:
\begin{equation}
    H=\sqrt{(k\left(T_{\rm mid0}\left(\frac{r}{150\ \rm au}\right)^{q}\right)R^3)/(\mu m_H GM_*)}
\end{equation} A hydrostatic equilibrium calculation is performed at each radius to derive the gas volume density at each height above the midplane, based on the temperature and surface density structure. CO is spread throughout the disc assuming a constant abundance. In regions with $N_{H_2}<1.3\times10^{21}$ cm$^{-2}$, where $N_{H_2}$ is the column density of H$_2$, CO is assumed to be photo-dissociated \citep{Qi2011} and the abundance in this region is dropped by eight orders of magnitude. Similarly, the CO abundance is dropped by a factor of five in regions with $T_{\rm gas}<19$ K to account for freeze out. We assume isotope abundances of $^{18}$O/$^{16}$O=557 and $^{13}$C/$^{12}$C=69 \citep{Wilson1999} when modeling $^{13}$CO and C$^{18}$O emission. 

Extended CO emission has been detected around IM Lup, at a distance where the disc gas temperatures would suggest that CO should be frozen out onto dust grains, indicating that photodesorption returns a substantial mass of CO to the gas phase \citep{cleeves16b,seifert21}. To accommodate this additional CO, we include CO in the outer disc following the prescription in \citet{flaherty20}: in the region of the disc where the gas temperature is below the CO freeze-out temperature, we include CO at a constant abundance in the region with $1.3\times10^{21}$ cm$^{-2}< N_{H_2} < 4.8 \times 10^{21}$ cm$^{-2}$. This prescription maintains CO freeze-out at the midplane, while allowing CO gas in the upper layers of the outer disc where high-energy photons can dislodge CO from ice mantles on dust grains. 

\citet{cleeves16b} find evidence for a factor of 20 depletion in CO abundance around IM Lup relative to the standard ISM value, while \citet{zhang19} find that the CO depletion is a factor of 100 in the inner disc ($<$100 au) and is a factor of 5 in the outer disc. We assume a uniform CO depletion of a factor of 20 across the entire disc in our models of the CO emission (i.e., CO/H$_2$ = 5$\times10^{-6}$), although we also consider models in which the CO has an abundance equal to the ISM value when fitting all three lines together. 

The velocity profile is assumed to be Keplerian, with additional corrections for the pressure support of the gas and the height above the midplane. The line profile is assumed to be a Gaussian whose width is set by thermal and non-thermal motions. We assume the non-thermal term is proportional to the local isothermal sound speed:
\begin{equation}
\Delta V = \sqrt{\left(2 k_B T(r,z)(1+\delta v_\mathrm{turb}^2)/m_{CO}\right)}%\sqrt{\left(2k_BT(r,z)/m_{CO}\right)(1+\delta v_{\rm turb}^2)}
\end{equation}
, where $\delta v_{\rm turb}$ is in units of c$_s$. Throughout much of this paper we assume that the non-thermal line broadening is associated with turbulence, although we cannot rule out non-Keplerian non-thermal motion below the spatial resolution limit. In section 4 we discuss such non-Keplerian non-thermal terms in the context of specific physical models (e.g., vertical motions associated with the VSI), and the effect of these types of motion on our turbulence measurement.

We assume M$_*$ = 1.1 M$_{\odot}$ \citep{alcala17,oberg21}. We use a distance of 158.5 pc from \textit{Gaia} DR2 measurements of the parallax \citep{bailer-jones18} and a gas mass of 0.175 M$_{\odot}$ \citep{pin18}. During each trial we vary $q$, $R_c$, $\delta v_{\rm turb}$, $T_{\rm atm0}$, $T_{\rm mid0}$, inclination ($i$), $R_{\rm in}$, systemic velocity ($v_{\rm sys}$), RA offset from phase centre ($x_{\rm off}$), Dec offset from phase centre ($y_{\rm off}$), and position angle (PA). For a given set of model parameters, and the resulting density, temperature, velocity structure, model images are generated by solving the radiative transfer equation, as described in \citet{rosenfeld13}. Images are generated at the same velocities as the data, with Hanning smoothing applied to the model images.

The posterior distributions for each parameter are estimated using the MCMC routine \texttt{EMCEE} \citep{foreman-mackey13}, based on the Affine-Invariant algorithm originally proposed in \citet{goodman10}. This methodology allows us to simultaneously constrain the temperature, density, and turbulence in the disk, accounting for any degeneracies between the parameters. The likelihood of a given model is calculated using the visibilities, with the model visibilities derived from model images using \texttt{galario} \citep{Tazzari2018}. 
\begin{equation}
    \ln{L} = -\frac{1}{2}\Sigma \frac{(V_{data,i}-V_{mod,i})^2}{\sigma_i^2}
\end{equation}
where the summation is performed over every $u$ and $v$ position and over every channel, and $V_{data,i}$ is a visibility point in the data at a given $u$ and $v$ and channel, while $V_{mod,i}$ is the visibility point at the same $u$, $v$, and channel derived from the model image. \texttt{Galario} has previously been applied to continuum images, and to extend it to spectral data we applied \texttt{galario} to each channel individually to derive the visibilities. The uncertainties in the data are estimated based on the calculated dispersion among baselines of similar distances in line-free channels. 

Absorption from the molecular cloud obscures some of the CO emission \citep{vankempen07,panic09} and when fitting to CO(2--1) we exclude channels with $v_{\rm LSR}$ between 4 and 6 km s$^{-1}$ (Figure~\ref{figure:spec}). Previous studies of turbulence using molecular line emission have focused on high SNR, un-contaminated targets, but expanding the sample requires the consideration of less ideal data, including objects with emission contaminated by cloud absorption similar to what is seen around IM Lup. Part of the goal of our study is to understand how this cloud absorption affects our ability to constrain turbulence in the disk around IM Lup. This absorption is not evident in $^{13}$CO or C$^{18}$O and we use the entire line profile when fitting these lines. Typical MCMC chains consist of 50 walkers and 1000 steps, with convergence on the final solution occurring within 300 steps. The first 500 steps are removed as burn-in, much longer than the typical auto-correlation time, which converges towards $\sim$60 steps towards the end of the chains.

\section{Results} \label{sec:results}

\begin{figure}%[ht]
\includegraphics[scale=.5]{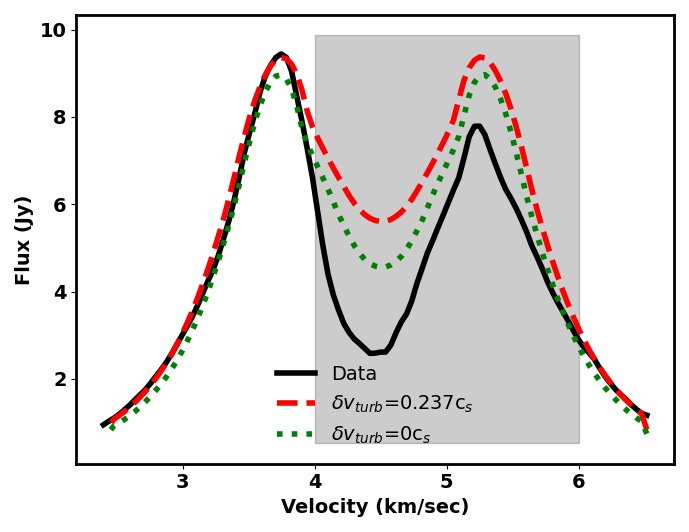}
\caption{CO(2--1) spectra of the disc around IM Lup (black line) and the median of the PDFs from the fiducial model (red dashed line). The region that is subject to absorption by the molecular cloud, and is excluded from the MCMC process, is marked with a grey band. Despite the limited spectral range, we are able to place strong constraints on the non-thermal linewidth ($\delta v_{\rm turb}$=0.237$^{+0.017}_{-0.012}$c$_s$), with a significantly better fit to the data than with zero turbulence (green dotted line). \label{figure:spec}}
\end{figure}

\begin{figure*}%[ht]
\center
\includegraphics[scale=.41]{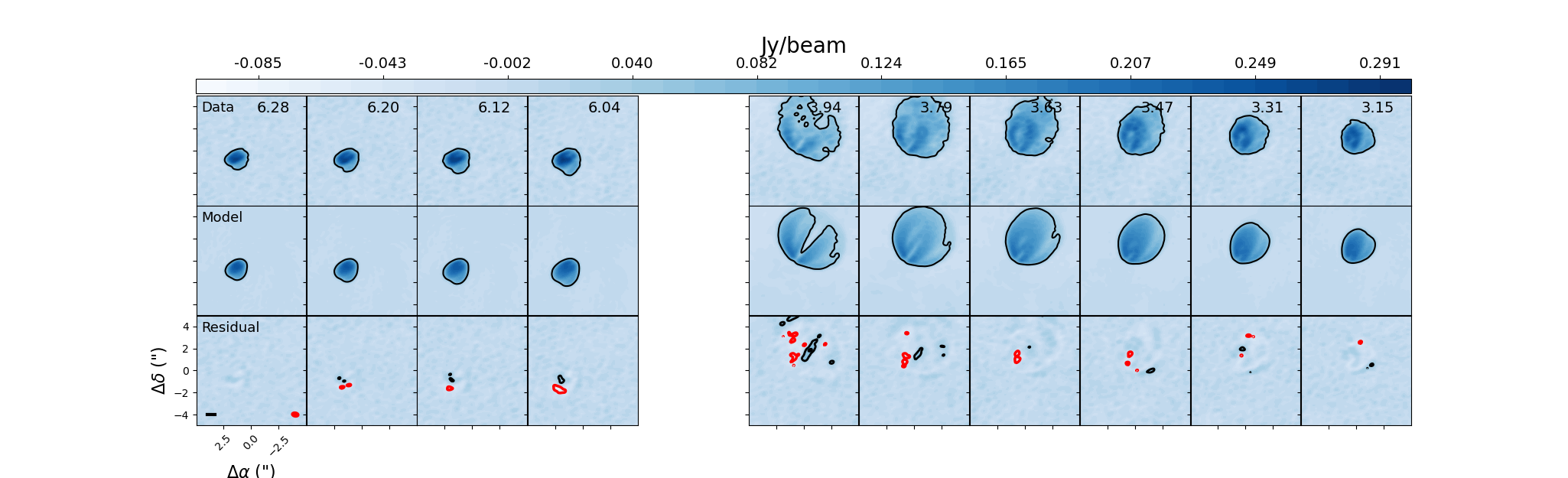}
\caption{Channel maps of the CO(2-1) data (top row), the fiducial model (middle row), and residuals (bottom row). In the data and model channel maps the contour indicates the 5$\sigma$ level ($\sigma$ = 9 mJy/beam). In the residual map, positive (black) and negative (red) contours are at levels in multiples of 5$\sigma$. The lower left panel includes a 100 au scale bar and the beam shape. The model reproduces much of the emission, although it does underestimate the emission near the midplane.
\label{figure:imdiff}}
\end{figure*}

With our fiducial model, fit to the CO(2--1) data, excluding the region with $v_{\rm LSR}$ between 4 and 6 km s$^{-1}$, we find turbulence is non-zero, at $\delta v_{\rm turb}$=0.237$^{+0.017}_{-0.012}$c$_s$ (Table~\ref{table:results}). Figure~\ref{figure:spec} shows the spectrum of the data and the model defined by the median of the posterior distribution functions (PDFs), while Figure~\ref{figure:imdiff} shows the residuals between the model and the data. The model is an excellent match to the data, with some small positive residuals towards the midplane of the disc. We find that we are able to constrain turbulence despite the exclusion of some of the channels. This is because each channel contains information about the turbulence, which we are able to utilize by modeling the full three-dimensional data set.

\begin{figure}%[ht]
\includegraphics[scale=.5]{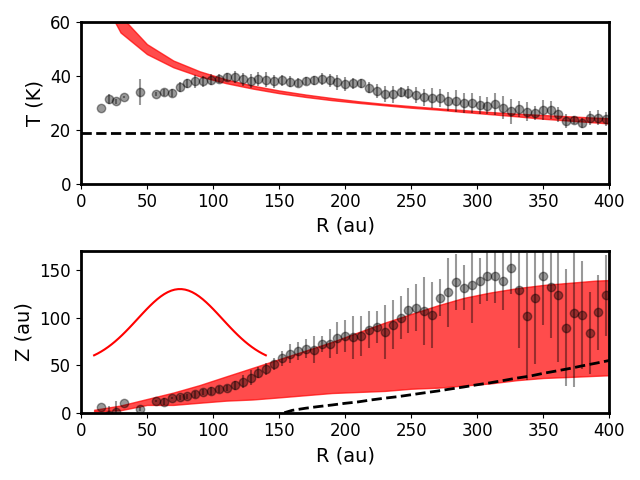}
\caption{Temperature (top panel) and height of the emitting region (bottom panel) vs radius for our fiducial model (red band) and as derived directly from the data by \citet{law21} (points). In the top panel the dashed line indicates a CO freeze-out temperature of 19 K, while in the bottom panel the dashed line indicates the location of the CO condensation front in our fiducial model. The extended emitting height for CO(2--1) comes from the fact that we see both the near and far side of the disc; the top of this region is seen on the near side of the disc, while the bottom is seen on the far side of the disc. The red Gaussian indicates the beam size in the data we analyze. \label{figure:Temp_Z_fiducial}}
\end{figure}
%We are somewhat lower than the Law et al. results, by about 20% throughout much of the outer disc. In general the intensity in our data seems to be 20% lower than for the MAPS data. 
%Oberg et al. 2021 list an integrated flux of 22.432+-.094 Jy km/s for CO(2-1)
%My integrated flux is 16.96 Jy km/s, 30% lower.  --> Did my self-calibration screw things up?? I don't do amplitude self-cal so that shouldn't be the issue. 
%Pinte et al. report a disc integrated flux of 24.1+-2.4 Jy km/s. 

Figure~\ref{figure:Temp_Z_fiducial} shows the temperature and height of the CO emitting regions as derived in our fiducial model, as compared to the height and temperature derived by \citet{law21}. The temperature and height of the CO emitting regions were derived by considering the distribution of temperatures and heights that contribute to the lines of sight that pass through a given radial bin; the regions in Figure~\ref{figure:Temp_Z_fiducial} show the 25th and 75th percentile points on this distribution at each radius, representing the boundaries within which the majority of the emission at a given radius arises. The breadth of this region represents the range of optical depths for the lines of sight at a given radius, as well as the front- and back-sides of the disk. As seen in the bottom-middle and bottom-right panels of Figure~\ref{figure:channel}, while the majority of the emission is optically thick and comes from a consistent $Z$ above the midplane, some emission arises from deeper in the disk. We are consistent with the \citet{law21} results, although we derive a higher temperature in the inner disc, where the Law et al. measurements are biased by beam dilution. A comparison of temperatures derived with these different data sets will also be subject to the 10-20\% uncertainty in the amplitude calibration \citep{Butler2012}. Given the temperatures of the CO emitting region, the turbulence constraint ($\delta v_{\rm turb}$=0.237$^{+0.017}_{-0.012}$c$_s$, c$_s$=$\sqrt{2 k T/\mu m_h}$) corresponds to 87 m s$^{-1}$ at 100 au and 65 m s$^{-1}$ at 500 au \footnote{Corresponding $\alpha$ values are discussed in Sections \ref{sec:scattered_light} and \ref{sec:discussion}, with the caveat that not all $\alpha$ measurements are the same.}. These velocities are consistent with the models of \citet{cleeves16b}, which find that turbulence is weaker than 200 m s$^{-1}$ (0.5 -- 0.7 c$_s$). Our results are slightly weaker than the (0.4--0.6)c$_s$ turbulence derived by \citet{paneque-carreno2023} using CN emission. Unlike our simultaneous modeling of the thermal and non-thermal component, \citet{paneque-carreno2023} assume a gas temperature from the intensity of optically thick CO or $^{13}$CO and compare this to the broadening observed in CN.

While the statistical uncertainty on the turbulence is small within the fiducial model, systematic effects amplify this uncertainty. Accounting for amplitude calibration uncertainty, by modeling the emission with a $\pm$20\% offset in total flux (models {\it High sys} and {\it Low sys} in Table~\ref{table:results}), we find a statistically significant difference in $T_{\rm atm0}$ (29.2 -- 39.1 K) and $\delta v_{\rm turb}$ (0.18 -- 0.28 c$_s$). Increasing $Z_q$ from 2 times the pressure scale height to 4 times the pressure scale height ({\it $Z_q=4H$} in Table~\ref{table:results}) also results in a turbulence that is statistically significantly different from the fiducial model ($\delta v_{\rm turb}$ = 0.30$^{+0.01}_{-0.02}$ c$_s$) but also does not change our overall interpretation that there is non-zero turbulence around IM Lup. The stellar mass also influences the velocity pattern, and including $M_*$ as a free parameter ({\it Stellar Mass} in Table~\ref{table:results}) does not substantially change the turbulence ($\delta v_{\rm turb}$ = 0.245$^{+0.010}_{-0.018}$ c$_s$), while finding a stellar mass ($M_*=1.088^{+0.007}_{-0.006}$ M$_{\odot}$) that is consistent with analysis of the velocity profile \citep[$M_*\approx 1.02$ M$_{\odot}$][]{Lodato2023}. 

Previous work has found that the ratio of the peak flux to the flux at line centre varies with the turbulence level, with smaller peak-to-trough ratios associated with stronger turbulence \citep{simon15,pinte2022}. While the cloud contamination prevents us from accurately measuring the peak-to-trough ratio, we are still able to robustly measure the turbulence, albeit with larger systematic uncertainties. In the next two subsections we examine in detail the influence of midplane temperature and CO abundance on turbulence, but the general result still holds: we find robust evidence for non-zero turbulence in the disc around IM Lup, at the level of (0.18 -- 0.30)c$_s$.

\subsection{Midplane Temperature}
Our result predicts a CO snowline (T$_{\rm gas}$=19 K) at the midplane at 154$^{+15}_{-12}$ au, while \citet{qi19} find an upper limit on the CO snowline radial location of 59 au based on the innermost edge of the N$_2$H$^+$ emission. \citet{seifert21}, building on models from \citet{cleeves16b}, also find a CO snow line that is smaller than our model predictions, while \citet{zhang21} put the midplane CO snowline at 15$\pm$5 au. The CO emission studied here is optically thick and does not directly trace the midplane temperature. Even the emission from the far side of the disc arises from slightly above the midplane (Figure~\ref{figure:Temp_Z_fiducial}). In this way, the midplane temperature does not directly affect the CO emission. But the midplane temperature does indirectly affect our models through the gas pressure scale height, which roughly scales with the square root of the midplane temperature. Since the height of the emitting region is constrained by the data, the midplane temperature is indirectly constrained and the large height of the CO emitting surface \citep{pin18,law21} results in a large midplane temperature, which pushes the CO snowline in our model to large radii. The potentially over-estimated midplane temperature may also reflect missing features in the parametric model, such as a jump in the midplane temperature beyond the outer edge of the dust disc \citep{Cleeves16a}, or an increasing CO abundance with radius \citep{zhang21}.

\citet{flaherty18} find that the choice of midplane temperature can bias the turbulence measurement, with an over-estimate of the midplane temperature leading to an underestimate of the turbulence. To understand this effect in the disc around IM Lup, we run a trial with T$_{\rm mid0}$ set such that the snow line occurs at 59 au (i.e., T$_{\rm mid 0}$ = 19$\times$(59/150)$^{-q}$), and a second trial where the midplane snowline is at 19 au. As with the fiducial model, we find non-zero turbulence, at a level of $\delta v_{\rm turb}$=0.25$^{+0.01}_{-0.02}$ c$_s$ and 0.256$^{+0.018}_{-0.010}$ c$_s$ for the models with the snowline at 59 au and 15 au respectively({\it R$_{snow CO}=59$} and {\it R$_{snow CO}=15$} in Table~\ref{table:results}). This is consistent with the findings of \citet{flaherty18} that a lower midplane temperature is associated with stronger turbulence, although here the effect is at the level of the uncertainty in the turbulent velocity. This indicates that midplane temperature does not strongly affect our estimate of the turbulence.

\subsection{CO abundance}
\begin{figure*}
    \centering
    \includegraphics[scale=.36]{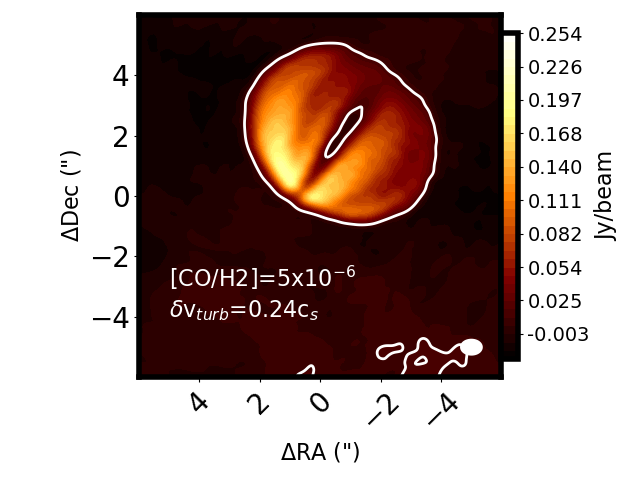}
    \includegraphics[scale=.36]{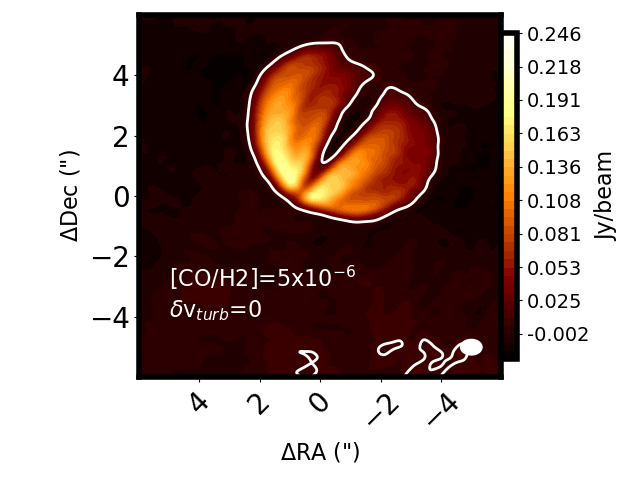}
    \includegraphics[scale=.36]{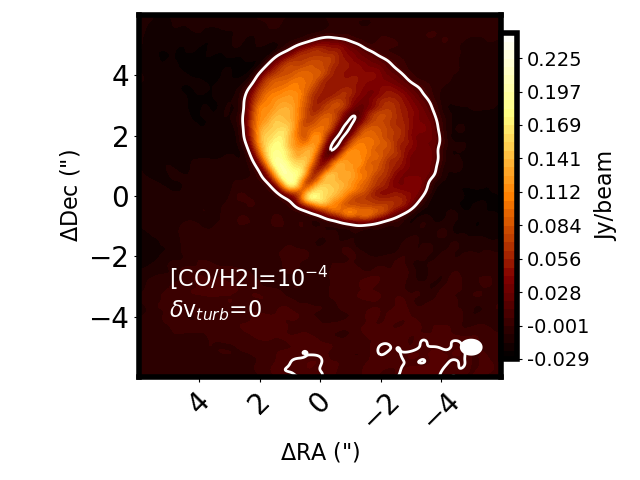}
    \includegraphics[scale=.36]{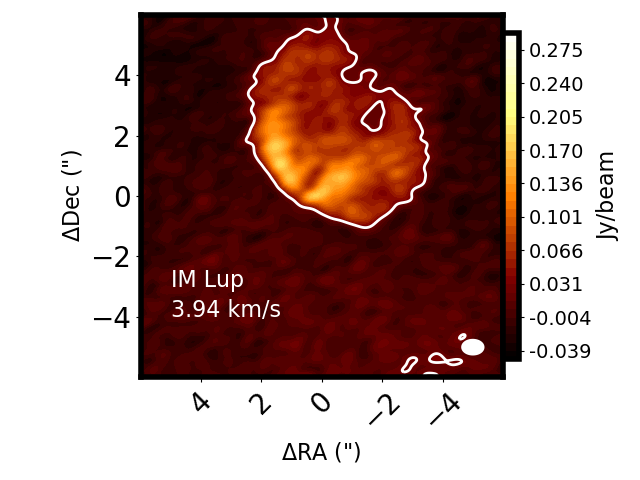}
    \includegraphics[scale=.36]{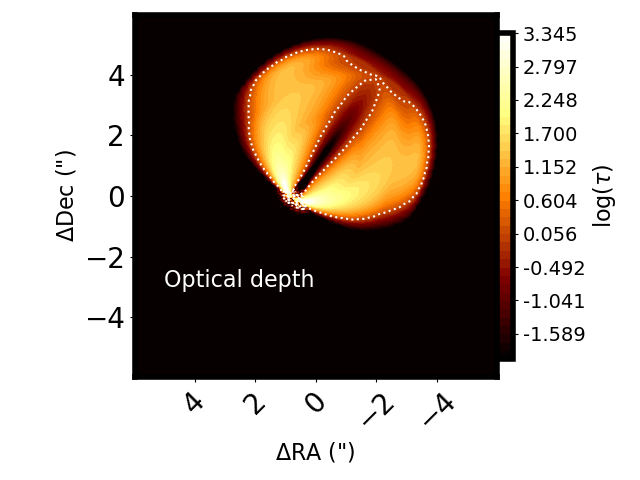}
    \includegraphics[scale=.36]{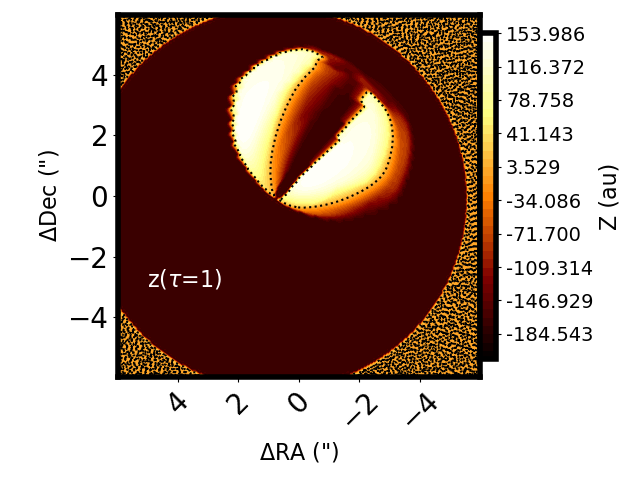}
    \caption{Top row: Individual channel maps for three different models, with varying levels of turbulence and CO abundance (the white contour indicates the 5$\sigma$ level). There is a degeneracy between turbulence and CO abundance, such that low turbulence can be matched with a higher CO abundance. This can be seen in comparing the morphology of the left and right panels, which is similar, despite the differences in abundance and turbulence. 
    Bottom Row: IM Lup data, followed by maps of optical depth (dotted line corresponds to $\tau=1$) and height of the $\tau=1$ surface (dotted line corresponds to z=0) for the fiducial model ([CO/H$_2$]=5$\times10^{-6}$, $\delta v_{\rm turb}=0.24$c$_s$). CO(2--1) is sensitive to the CO abundance, despite being mostly optically thick, because there are lines of sight that are optically thin (i.e., outside of the region marked by the dotted lines in the central panel) that reach deep into the disc. }
    \label{figure:channel}
\end{figure*}

While midplane temperature does not substantially affect our measurement of non-zero turbulence, we do find that the global CO abundance plays an important role in the measured turbulence level. When running a trial with no CO depletion (i.e., CO/H$_2$=10$^{-4}$) outside of CO freeze-out, we find $\delta v_{\rm turb}<$0.03c$_s$, much weaker than when CO depletion is included. This anti-correlation between turbulence and CO abundance is similar to that found when using more detailed chemical models of CO depletion \citep{Yu2017}. The connection between turbulence and CO abundance exists because both contribute to emission in the same regions of the channel maps. Figure~\ref{figure:channel} shows an individual channel from models with (1) low CO abundance and high turbulence, (2) low CO abundance and no turbulence, and (3) high CO abundance with no turbulence. Both the low abundance and high turbulence model (upper left panel) as well as the high abundance and low turbulence model (upper right panel) generate a broadening of the emission in the channel maps. This broadening is not present in the low abundance, zero turbulence model image (upper centre panel). This arises because even though CO is, overall, highly optically thick, there are lines of sight that are optically thin, making the emission along these lines of sight sensitive to the abundance. These lines of sight are the same lines of sight that are most strongly influenced by the turbulence, as they come from regions that have bulk motions at the edge of the velocity channel, which get brought into the channel by turbulent motion.  

%channel 65 has v=3.94

This effect is larger than that seen around TW Hya, where increasing the CO abundance by a factor of 10 led to a change in the upper limit from $<$0.13c$_s$ to $<$0.08c$_s$ \citep{flaherty18}. This may be because the cloud contamination for IM Lup excludes the central channels. %As shown previously \citep{simon15}, the peak-to-trough ratio is sensitive to the turbulence level, and without this information we are more susceptible to the degeneracy with CO abundance. 
As shown previously \citep{simon15}, the peak-to-trough ratio is sensitive to the turbulence level, through its impact on the central velocity channel, and without this information we are more susceptible to the degeneracy with CO abundance. 

\begin{figure*}%[ht]
\includegraphics[scale=.5]{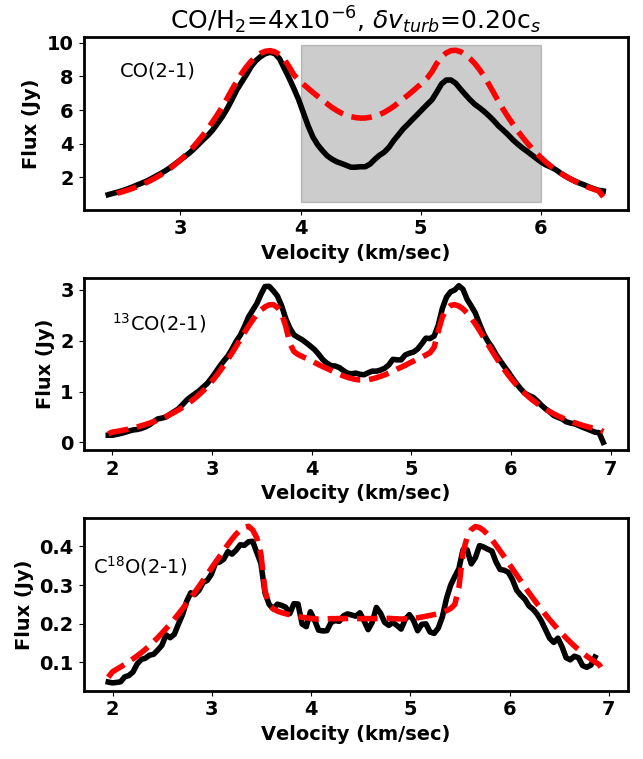}
\includegraphics[scale=.5]{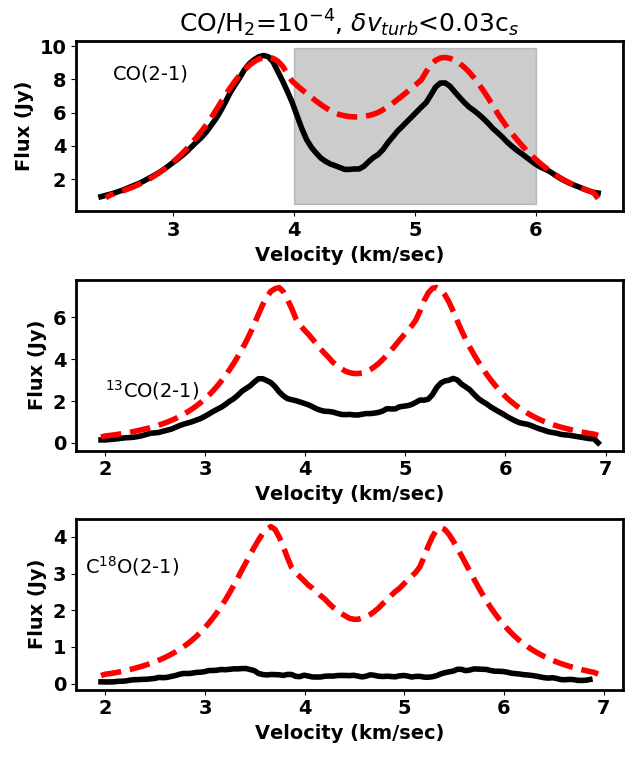}
\caption{CO(2--1) spectrum (top panels), $^{13}$CO(2--1) spectrum (middle panels) and C$^{18}$O(2--1) spectrum (bottom panels) of the disc around IM Lup (black line) and the median of the PDFs from the multi-line fit (red dashed line). We are able to fit all three spectral lines with non-zero turbulence and with significant CO depletion (left panels), but significantly overestimate the $^{13}$CO and C$^{18}$O flux when assuming no CO depletion (right panels. \label{figure:spec_all}}
\end{figure*}

\begin{figure}%[ht]
\includegraphics[scale=.4]{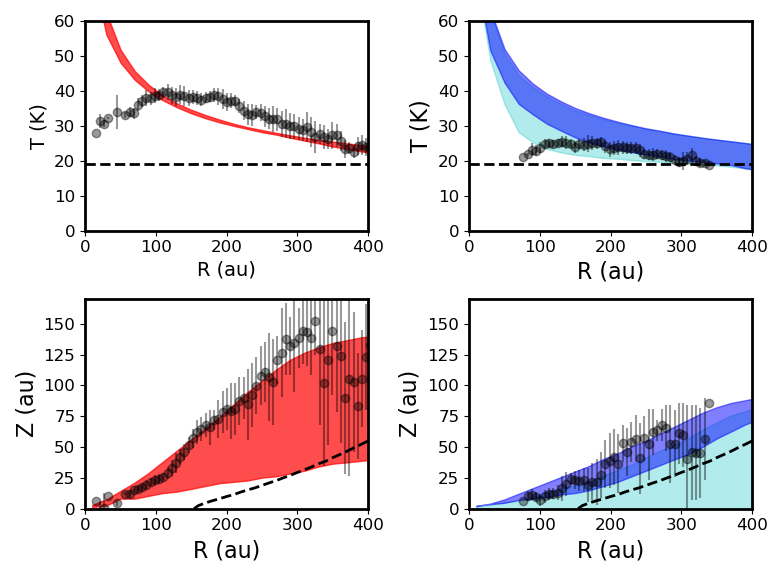}
\caption{Temperature (top panel) and height of the emitting region (bottom panel) vs radius for our multi-line model fit. The left panels show CO(2--1) in red, while the right panels show $^{13}$CO(2--1) in blue and C$^{18}$O(2--1) in cyan, with temperatures and emitting heights derived by \citet{law21} for CO(2--1) and $^{13}$CO(2--1) shown as circles. Our multi-line fit reproduces the temperature and height of the emitting region for these isotopologues. \label{figure:Temp_Z_all}}
\end{figure}

Without access to the central channels, other data is needed to break the degeneracy between turbulence and CO abundance. The CO(2--1) emission by itself cannot constrain the CO abundance, but when modeled in concert with more optically thin emission lines like $^{13}$CO(2--1) and C$^{18}$O(2--1) we can constrain the turbulence and CO abundance simultaneously. To this end, we simultaneously model CO(2--1), $^{13}$CO(2--1) and C$^{18}$O(2--1), while adding the CO abundance as a free parameter. For simplicity we assume a single turbulence scaling relative to the local sound speed, although models of MRI \citep[e.g.][]{simon15,simon18} and VSI \citep{flock17} indicate that turbulence likely varies with height within the disc, beyond a simple scaling with the local sound speed. We use a common set of physical parameters to create a separate set of model visibilities for all three isotopologues and sum the log-likehoods from all three ($\ln L_{tot} = \ln L_{CO}+\ln L_{13CO}+\ln L_{C18O}$). No additional weighting is applied to the log-likelihood values from each isotopologue, although the result is likely weighted toward CO(2-1) given its higher SNR. Results are shown in Table~\ref{table:results} ({\it Multi-line}) and Figure~\ref{figure:spec_all}; we find non-zero turbulence ($\delta v_{\rm turb}$=0.20$^{+0.01}_{-0.02}$ c$_s$), with a strong CO depletion (CO/H$_2$ = 4.1$^{+0.2}_{-0.3}\times10^{-6}$). The temperatures and heights of the emitting regions for these molecules are consistent with those values derived by \citet{law21} (Figure~\ref{figure:Temp_Z_all}). The CO abundance derived here is consistent with the factor of 20 depletion (i.e., CO/H$_2$ = 5$\times10^{-6}$) derived by \citet{cleeves16b} and the factor of 10 -- 100 depletion derived by \citet{zhang21} for this system. 

For simplicity we assumed a constant CO depletion across the disc, although observations and models suggest that the CO depletion may vary with radius \citep{zhang19,zhang21}, with height \citep{Krijt2020}, and with isotopologue \citep{Miotello2014}. \citet{zhang21} find that the CO depletion is a factor of 100 at radii less than 100 au, while the CO depletion is only a factor of 5 at larger radii, bracketing the CO depletion we find, suggesting that our result is best interpreted as an intensity-weighted average CO depletion throughout the disk. Given the degeneracy between CO abundance and turbulence, a spatially-varying CO abundance will most strongly affect efforts to constrain spatially-varying turbulence, which are beyond the scope of this paper. 

The multi-line fit converges on a turbulence level that is smaller than the single line fit. This may be due to a decrease in the turbulence level towards the midplane; if the $^{13}$CO(2--1) and C$^{18}$O(2--1) lines are probing regions with weaker turbulence (relative to the local sound speed) then fitting these lines will drag down the average turbulence level. As noted above, this behavior is expected from both MRI and VSI. Despite the similar vertical behavior, VSI may generate unique and detectable radial structure in the channel maps, as discussed in more detail below. Modeling the $^{13}$CO(2--1) line by itself ({\it $^{13}CO$ only} in Table~\ref{table:results}), or the C$^{18}$O(2-1) line by itself ({\it $C^{18}O$ only} in Table~\ref{table:results}), results in turbulence levels consistent with that derived from CO(2--1) by itself (0.237$^{+0.017}_{-0.012}$ c$_s$ for CO(2--1), 0.23$\pm0.04$ c$_s$ for $^{13}$CO(2--1), 0.25$^{+0.07}_{-0.15}$ for C$^{18}$O(2-1)), albeit with larger uncertainties even before accounting for systematic effects. Further analysis is needed to fully characterize any evidence for a vertical gradient in the turbulence from the line emission.

\subsection{Scattered light \label{sec:scattered_light}}
We can also characterize the turbulence level using the scattered light images of \citet{avenhaus18}. The height of the scattered light surface is related to the dust scale height, which in turn is set by turbulence. The $\alpha$ value is related to the dust scale height $H_d$, and the gas pressure scale height, $H$, via \citep[e.g.,][]{dullemond18}\footnote{This formula assumes that the dust-to-gas ratio is not large, otherwise the dust momentum itself acts to dampen the turbulence slightly, which in turn reduces the dust scale height \citep{Lin2019,Xu2022}.}:
\begin{equation}
    H_d=H_g(1+\psi^{-2})^{-1/2}
\end{equation}
where $\psi=\sqrt{\alpha/St}$. The Stokes number, $St$, is defined as \begin{equation}
    St = \frac{a \rho_s \pi}{2\Sigma_g}
\end{equation}
where $a$ is the particle size, $\rho_s$ is the bulk density of the material (=2 g cm$^{-2}$), and $\Sigma_g$ is the gas density. 

%\begin{equation}
%    \frac{\alpha}{St} = [(H/H_d)^2-1]^{-1}
%\end{equation}
%where $St$ is the Stokes number. The Stokes number is defined as:
%\begin{equation}
%    St = \frac{a \rho_s \pi}{2\Sigma_g}
%\end{equation}
%where $a$ is the particle size, $\rho_s$ is the bulk density of the %material (=2 g cm$^{-2}$), and $\Sigma_g$ is the gas density. 

To estimate $H_g/H_d$ from the SPHERE observations we generate synthetic scattered light images using \texttt{RADMC-3D} \citep{dullemond12}. We utilize the same density and temperature structure as in the fiducial CO model ({\it Fiducial} in Table~\ref{table:results}). Initially we consider the dust as a population of 0.1$\mu$m amorphous silicate grains spread uniformly throughout the radial extent of the disk. The gas to dust mass ratio is assumed to be 100, and the volume density of the dust has a Gaussian profile in the vertical direction with a scale height that is a fraction of the gas scale height (i.e., H$_d$/H is constant throughout the disk), where the gas scale height is approximated by H$_g$=c$_s$/$\Omega$, where c$_s$ is defined by the midplane temperature. Scattered light images are generated assuming isotropic scattering at the wavelength of the observations from \citet{avenhaus18}. The scattered light images are deprojected using \texttt{diskmap} \citep{stolker2016} which assumes that the height of the scattering surface follows a power law with radius (z$_{\rm scatter}$ = z$_0$(r/ 1 au)$^{a}$). Under the assumption of isotropic scattering, a deprojection that uses a combination of z$_0$ and $a$ that accurately reflects the height of the scattered light surface will produce emission that is axisymmetric about the center of the disk (i.e., radial profiles in any direction away from the center of the disk will be identical). For a given H$_d$/H$_g$, which we vary between 0.3 and 1, we search for a combination of z$_0$ and $a$ that minimizes the asymmetry of the emission, and exclude values of H$_d$/H$_g$ for which the combination of z$_0$ and $a$ that minimize the asymmetry are inconsistent with the height of the rings observed by \citet{avenhaus18}.

\begin{figure}%[ht]
\centering
\includegraphics[scale=.55]{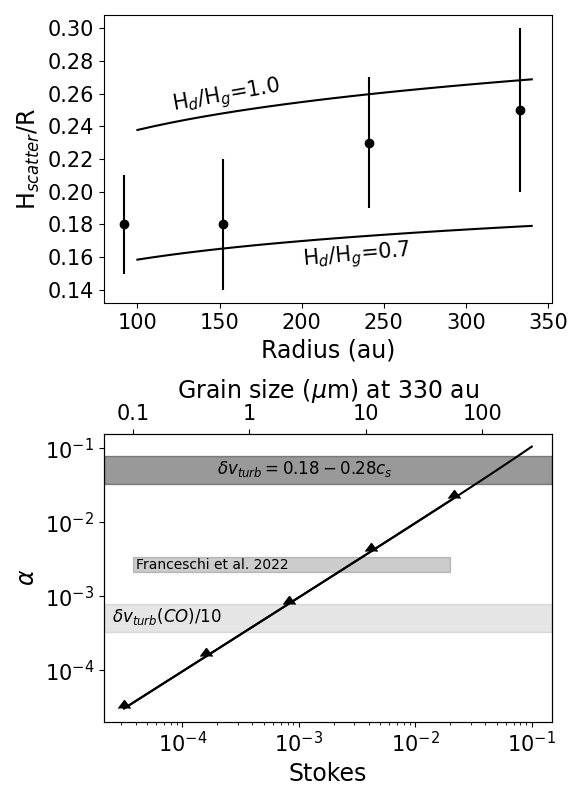}
\caption{(Top): Height of the scattered light surface, normalized to the radius, for the dust rings identified in the scattered light image from \citet{avenhaus18} (black dots), along with models with different dust scale heights. Ratios of dust to gas scale height between 0.7 and 1.0 match the height of the scattered light rings.
(Bottom): Constraints on $\alpha$ based on the dust scale height, as a function of the Stokes number. The dark grey band is the constraint on turbulence from the CO observations with CO depletion. The constraint on $\alpha$ from the dust scale height (solid line with arrows) is a lower limit given the asymptotic behavior of $\alpha$ as H$_d$/H$_g$ approaches 1, and is consistent with the results from \citet{Franceschi2022}, who simultaneously constrain $\alpha$ and the maximum grain size. These constraints on the dust settling are also consistent with a decrease in the turbulence between the surface layers probed by CO, and the midplane probed by dust settling. \label{figure:HR_alpha}}
\end{figure}

We find that $H_d/H_g\approx0.7 - 1$ (Figure~\ref{figure:HR_alpha}), indicating that $\psi\gtrapprox1$. While our methodology includes some limiting assumptions, they are consistent with a large H$_d$/H$_g$. Dust coagulation models, accounting for growth, fragmentation, and transport, predict a continuous distribution of grain sizes \citep{Birnstiel2012}. Extending our modeling to include a mix of 0.1 and 100$\mu$m grains, with the large grains have a scale height ten times smaller than the small dust grains, reduces the height of the scattered light surface for a given H$_d$/H$_g$. As a result, including a more realistic grain size distribution would push our result towards higher H$_d$/H$_g$, making our lower limit of 0.7 a conservative estimate.

%, while varying the scale height of the dust\footnote{Dust coagulation models, accounting for growth, fragmentation, and transport, predict a continuous distribution of grain sizes \citep{Birnstiel2012}. For simplicity, we assume a simple grain population composed of 0.1 and 100$\micron$ grains, and assume that the larger grains, which do not contribute significantly to the scattered light emission, have a scale height 10 times smaller than the small dust grains.}. We deproject the generated images to determine the height of the scattering surface, which is then compared to the results from \citet{avenhaus18}.

Given the asymptotic behavior of $\alpha$ as H$_d$/H$_g$ approaches 1 ($\alpha\to \infty$ as H$_d$/H$_g$ $\to$1), taking $H_d/H_g=0.7$ provides a lower limit on $\alpha$, which itself scales linearly with the Stokes number. For a broad range of Stokes numbers, the lower limit on $\alpha$ is consistent with the turbulence constraint from the CO emission. 

When dust is highly coupled to the gas ($St\to0$ or $\alpha\to \infty$) the scale height of the dust will depend on the turbulence throughout the vertical extent of the disc. When dust is not perfectly coupled to the gas, dust settling is more strongly dependent on the turbulence at the midplane \citep{ciesla07}, even though we are observing light scattering off of small dust grains in the atmosphere of the disk. CO, on the other hand, traces turbulence in the surface layers (Figure~\ref{figure:Temp_Z_fiducial}). Models of MRI and VSI indicate that turbulent velocities can decrease toward the midplane by up to a factor of ten \citep{simon15,flock17}. A factor of ten lower turbulence is indicated in Figure~\ref{figure:HR_alpha}, and the consistency between this velocity and the scattered light image depends on the size of the dust grains in the disk atmosphere that are responsible for the scattering. \citet{qi19}, in modeling the SED of IM Lup, find a maximum grain size of 3$\micron$ for the disc atmosphere, while \citet{Tazaki2023} find that the colour and polarization of scattered light images are consistent with fractal dust aggregates larger than 2$\micron$. \citet{Franceschi2022} use the scattered light emission in combination with the continuum emission from ALMA to constrain $\alpha$ and the maximum grain size, finding results consistent with the level of turbulence we derive from CO, under the assumption that turbulence decreases towards the midplane. %Assuming we are seeing micron-sized grains in the atmosphere of the disc with the scattered light image, the constraint on $\alpha$ from \citet{Franceschi2022} is consistent with our CO observations, assuming a decrease in the turbulent motion between the surface layer and the midplane. 

\begin{deluxetable}{cccccccccccc}
\rotate
%\caption{MCMC trials}
\tabletypesize{\scriptsize}
\tablewidth{0pt}
\tablehead{\colhead{Model} & \colhead{$q$} & \colhead{$\log R_c (\rm au)$} & \colhead{$\delta v_{\rm turb}$ (c$_s$)} & \colhead{$T_{\rm atm0}$ (K)} & \colhead{$T_{\rm mid0}$ (K)} & \colhead{$i$ ($^{\circ}$)} & \colhead{$R_{\rm in}$ (au)} & \colhead{$v_{\rm sys}$ (km s$^{-1}$)} & \colhead{$x_{\rm off}$ ($\arcsec$)} & \colhead{$y_{\rm off}$ ($\arcsec$)} & \colhead{PA ($^{\circ}$)}\label{table:results}}
\startdata
{\it Fiducial} & -0.355$^{+0.006}_{-0.007}$ & 2.396$^{+0.017}_{-0.030}$ & 0.237$^{+0.017}_{-0.012}$ & 34.3$^{+0.3}_{-0.2}$ & 19.2$^{+0.6}_{-0.5}$ & 53.3$^{+0.3}_{-0.2}$ & $<$7 & 4.508$\pm0.002$ & 0.705$^{+0.016}_{-0.14}$ & -0.036$^{+0.011}_{-0.006}$ & 144.9$^{+0.3}_{-0.5}$\\ %co21_lowabund
{\it High sys} & -0.292$^{+0.007}_{-0.006}$ & 2.317$^{+0.011}_{-0.010}$ & 0.184$^{+0.011}_{-0.025}$ & 39.1$^{+0.3}_{-0.2}$ & 18.6$^{+0.4}_{-0.8}$ & 53.9$\pm0.3$ & $<9$ & 4.484$^{+0.002}_{-0.005}$ & 0.747$^{+0.004}_{-0.005}$ & -0.041$^{+0.008}_{-0.004}$ & 143.2$^{+0.5}_{-0.1}$\\%co21_lowabund_highsys
{\it Low sys} & -0.361$^{+0.005}_{-0.006}$ & 2.380$^{+0.024}_{-0.019}$ & 0.279$^{+0.009}_{-0.011}$ & 29.2$^{+0.1}_{-0.2}$ & 15.4$^{+0.5}_{-0.6}$ & 52.9$^{+0.3}_{-0.2}$ & $<$13 & 4.513$\pm0.002$ & 0.737$^{+0.010}_{-0.008}$ & -0.034$\pm0.004$ & 144.0$^{+0.3}_{-0.5}$\\ %co21_lowabund_lowsys
%%no turb & -0.329$\pm0.004$ & 2.414$^{+0.01}_{-0.009}$ & 0\tablenotemark{c} & 35.3$\pm0.1$ & 19.5$^{+0.3}_{-0.4}$ & 54.9$\pm0.1$ & $<$27 & 4.495$^{+0.001}_{-0.002}$ & 0.737$\pm0.006$ & -0.047$^{+0.003}_{-0.002}$ & 144.2$^{+0.2}_{-0.1}$\\ %co21_lowabund_noturb
{\it Z$_q$=4H} & -0.265$^{+0.006}_{-0.004}$ & 2.377$^{+0.011}_{-0.010}$ & 0.295$^{+0.013}_{-0.010}$  & 39.2$\pm0.2$ & 18.5$^{+0.3}_{-0.2}$ & 52.4$^{+0.3}_{-0.2}$ & $<$10 & 4.514$^{+0.002}_{-0.001}$ & 0.699$\pm0.004$ & -0.047$^{+0.004}_{-0.002}$ & 144.4$\pm0.2$\\ %co21_lowabund_highZq
{\it Stellar Mass\tablenotemark{a}} & -0.342$\pm0.007$ & 2.330$^{+0.014}_{-0.009}$ & 0.245$^{+0.01}_{-0.018}$ & 34.37$^{+0.15}_{-0.26}$ & 19.1$^{+0.4}_{-0.6}$ & 52.6$^{+0.3}_{0.2}$ & $<$8 & 4.491$^{+0.003}_{-0.004}$ & 0.687$\pm$0.004 & -0.030$\pm0.004$ & 144.67$^{+0.21}_{-0.17}$\\ %co21_lowabund_Mstar
{\it R$_{\rm snow,CO}$=19} & -0.342$^{+0.006}_{-0.005}$ & 2.399$^{+0.016}_{-0.010}$ & 0.250$^{+0.008}_{-0.018}$ & 34.4$\pm0.2$ & 13.0\tablenotemark{b} & 52.9$^{+0.1}_{-0.2}$ & $<$39 & 4.507$\pm0.002$ & 0.727$^{+0.005}_{-0.004}$ & -0.034$^{+0.004}_{-0.003}$ & 144.0$^{+0.2}_{-0.1}$\\ %co21_lowabund_lowTmid
{\it R$_{\rm snow,CO}=59$} & -0.375$^{+0.008}_{-0.022}$ & 2.406$^{+0.026}_{-0.021}$ & 0.256$^{+0.018}_{-0.010}$ & 34.5$\pm0.3$ & 8.0\tablenotemark{c} & 52.5$^{+0.3}_{-0.2}$ & $<$16 & 4.493$^{+0.004}_{-0.002}$ & 0.752$^{+0.007}_{-0.010}$ & -0.012$^{+0.007}_{-0.004}$ & 144.1$\pm0.4$ \\ %co21_lowabund_vlowTmid
{\it Multi-line\tablenotemark{d}} & -0.348$^{+0.011}_{-0.009}$ & 2.525$^{+0.035}_{-0.067}$ & 0.201$^{+0.026}_{-0.022}$ & 35.0$^{+0.4}_{-0.5}$ & 18.0$^{+0.4}_{-0.9}$ & 53.9$^{+0.3}_{-0.4}$ & 14.2$^{+4.2}_{-1.2}$ & 4.511$^{+0.004}_{-0.005}$ & 0.699$^{+0.009}_{-0.013}$ & -0.047$^{+0.009}_{-0.005}$ & 145.0$^{+0.2}_{-0.5}$\\ %coAll
{\it Disc self-gravity\tablenotemark{e}} & -0.359$^{+0.006}_{-0.005}$ & 2.368$^{+0.012}_{-0.013}$ & 0.246$^{+0.019}_{-0.018}$ & 35.0$^{+0.4}_{-0.2}$ & 18.3$\pm0.6$ & 50.4$\pm0.3$ & $<$6 & 4.529$\pm0.003$ & 0.728$^{+0.004}_{-0.005}$ & -0.026$^{+0.006}_{-0.005}$ & 144.1$^{+0.2}_{-0.1}$\\ %co21_lowabund_sg
\hline
{\it High CO/H$_2$\tablenotemark{f}} & -0.388$^{+0.006}_{-0.008}$ & 2.334$^{+0.010}_{-0.008}$ & $<$0.03 & 34.5$^{+0.2}_{-0.1}$ & 17.7$^{+0.4}_{-0.6}$ & 52.7$\pm0.1$ & $<$7 & 4.511$^{+0.002}_{-0.001}$ & 0.706$^{+0.006}_{-0.005}$ & -0.027$^{+0.009}_{-0.003}$ & 144.2$\pm0.2$ \\ %co21_hpcc3
\hline
{\it $^{13}$CO only} & -0.10$^{+0.02}_{-0.03}$ & 2.41$^{+0.04}_{0.03}$ & 0.23$\pm0.04$ & 29.5$^{+1.1}_{-0.9}$ & 13.2$^{+1.1}_{-1.2}$ & 56.1$^{+0.2}_{-0.4}$ & $<13$ & 4.522$^{+0.005}_{-0.004}$ & 0.696$^{+0.013}_{-0.014}$ & 0.017$^{+0.013}_{-0.008}$ & 144.2$^{+0.5}_{-0.3}$ \\%13co21
{\it C$^{18}$O only} & -1.06$^{+0.11}_{-0.07}$ & 2.80$\pm0.04$ & 0.25$^{+0.07}_{-0.15}$  & 88$^{+12}_{-11}$ & 23$\pm2$ & 54.4$^{+0.8}_{-0.7}$ & $<62$ & 4.54$^{+0.01}_{-0.02}$ & 0.704$\pm0.012$ & -0.014$^{+0.010}_{-0.013}$ & 144.5$^{+0.6}_{-0.7}$ \\ %c18o21

\enddata
\tablenotetext{a}{Stellar mass was included as a free parameter: M$_*$=1.088$^{+0.007}_{-0.006}$ M$_{\odot}$}
\vspace{-3mm}
\tablenotetext{b}{T$_{\rm mid0} = 19~{\rm K}\times(59/150~{\rm au})^{-q}$,  putting the CO snowline at the midplane at 59 au.}
\tablenotetext{c}{T$_{\rm mid0} = 19~{\rm K}\times(15/150~{\rm au})^{-q}$,  putting the CO snowline at the midplane at 15 au.}
%\vspace{-2mm}
\tablenotetext{d}{Within the multi-line fit, the CO abundance is also a free parameter: log([CO/H$_2$]) = -5.39$^{+0.02}_{-0.03}$}
%\vspace{-2mm}
\tablenotetext{e}{Disc mass was included as a free parameter: M$_{\rm gas}$=0.14$^{+0.01}_{-0.06}$ M$_{\odot}$}
%\vspace{-2mm}
\tablenotetext{f}{While this model reproduces the CO(2-1) emission, it strongly over-predicts the $^{13}$CO and C$^{18}$O emission.}
%%\tablenotetext{c}{Turbulence fixed to zero in this trial.}
\end{deluxetable}

\section{Discussion: What is driving turbulence? } \label{sec:discussion}

In \citet{flaherty20} we found nonthermal gas motion, between 0.25 c$_s$ and 0.33 c$_s$, around DM Tau. Here we have examined IM Lup and found similar non-zero nonthermal gas motion at $\delta v_{\rm turb} = 0.18 - 0.30$ c$_s$. These results stand in contrast to the non-detections of turbulence around MWC 480, V4046 Sgr, TW Hya, and HD 163296, with upper limits ranging from $\delta v_{\rm turb}<0.15$ c$_s$ down to $\delta v_{\rm turb}<$0.05 c$_s$ \citep{flaherty18,Teague18,flaherty20} but are consistent with the finding of strong non-thermal motion around IM Lup by \citet{paneque-carreno2023}. In the context of an $\alpha$ disc model, with  $\alpha\sim(\delta v_{\rm turb}/c_s)^2$, and under the assumption that $\delta v_{\rm turb}/c_s$ is constant throughout the disc, our result corresponds to $\alpha=0.03 - 0.08$ around IM Lup, $\alpha = 0.06 - 0.10$ around DM Tau, and $\alpha\lesssim$0.01 around the other sources. \citet{Powell2022} examine the CO abundance in the context of gas diffusion and dust grain surface chemistry and find high gas diffusion, consistent with consistent with $\alpha=0.003-0.015$, around IM Lup and DM Tau and low diffusion, consistent with $\alpha\sim10^{-4}$, around HD 163296 and TW Hya. \citet{Bosman2021} find evidence for high turbulence ($\alpha>4\times10^{-3}$) in the inner 20 au of the disk around IM Lup, based on the dust obscuration of CO emission from the inner disk. These constraints, listed in Table~\ref{table:alpha_literature}, are in line with our results, with the caveat that $\alpha$ values directly derived from CO or CN emission are tracing the behavior in the upper layers of the outer disc rather than the midplane traced by CO diffusion, where turbulence is likely lower, or the inner disk traced by CO obscuration. 

\begin{deluxetable}{ccc}
    \tabletypesize{\footnotesize}
    \tablewidth{0pt}
    \tablehead{\colhead{$\alpha$} & \colhead{Method} & \colhead{Reference}\label{table:alpha_literature}}
    \startdata
    0.03 - 0.08 & CO/$^{13}$CO/C$^{18}$O J=2-1 & this paper\\
     & emission line modeling & \\
    %$>$??St & Dust settling & this paper\\
    0.003 - 0.015 & CO abundance with & \citet{Powell2022}\\
     & surface-grain chemistry & \\
    $>$4$\times10^{-3}$ & Dust obscuration of CO & \citet{Bosman2023}\\
     & at r$<$20 au & \\
    2.9$^{+0.5}_{-0.8}\times10^{-3}$ & Dust settling & \citet{Franceschi2022}\\
    0.16 - 0.36 & CN emission & \citet{paneque-carreno2023}\\
    \enddata
\end{deluxetable}

\begin{deluxetable}{cccccccccc}
    \tabletypesize{\footnotesize}
    \tablewidth{0pt}
    \rotate
    \tablehead{\colhead{Star} & \colhead{$\delta v_{\rm turb}$} & \colhead{M$_{*}$} & \colhead{M$_{\rm disk}$} & \colhead{Age} & \colhead{$\dot{M}$} & \colhead{L$_{\rm FUV}$} & \colhead{Ionization} & \colhead{R$_{CO}$} & \colhead{R$_{dust}$} \\
    \colhead{} & \colhead{} & \colhead{(M$_{\odot}$)} & \colhead{(M$_{\odot}$)} & \colhead{(Myr)} & \colhead{(M$_{\odot}$ yr$^{-1}$)} & \colhead{(L$_{\odot}$)} & \colhead{(s$^{-1}$)} & \colhead{(au)} & \colhead{(au)}\label{table:stellar_params}}
    \startdata
    HD 163296 & $<$0.05 c$_s$ (1) & 2.3 & 0.09 (2) & 5 & 5$\times10^{-7}$ (3) & 3.21 -- 5.58 (4) & $\gtrsim10^{-18}$ (5) & 478$\pm$5 (6) & $\sim$540 au (7)\\
    TW Hya & $<$0.08 c$_s$ (8) & 0.6 & 0.05 (9) & 10 -- 12 & 2$\times10^{-9}$ (10, 11, 12) & 7$\times10^{-3}$ (13) & $\lesssim 10^{-19}$ (14) & 184$\pm$4 (6) & $\approx$150 au (15)\\
    MWC 480 & $<$0.08 c$_s$ (16) & 1.85 (17) & 0.046 (16) & 7 -- 8 & 5.3$\times10^{-7}$ (18, 19, 20) & \ldots & \ldots & 595 (21) & $\le$350 au (22)\\
    V4046 Sgr & $<$0.12 c$_s$ (16) & 0.9, 0.9 & 0.09 (23) & 12 -- 23 (24, 25, 26) & 1.3$\times10^{-8}$ (27) & 10$^{-2}$ (13) & \ldots & 362$\pm$39 (6) & $\approx$250 au (28)\\
    DM Tau & (0.25--0.33) c$_s$ (16) & 0.54 (29) & 0.04 (30, 31) & 1 -- 5 & 2.9$\times10^{-9}$ (13, 32) & 3$\times10^{-3}$ (33, 13) & \ldots & 876$\pm$23 (6) & 58 au (34)\\
    IM Lup & (0.18--0.3) c$_s$ (35) & 1.1 & 0.175 & 1 (36, 37) & 10$^{-8}$ (38) & $\sim$8$\times10^{-3}$ (38) & $\gtrsim10^{-17}$ (39) & 803$\pm$9 (6) & $>$700 au (29)\\
    \enddata
    \tablecomments{References: (1) \citet{flaherty2017}, (2) \citet{isella2007}, (3) \citet{mendigutia2013}, (4) \citet{meeus12}, (5) \citet{aikawa21}, (6) \citet{long2022}, (7) \citet{Wisniewski2008}, (8) \citet{flaherty18}, (9) \citet{bergin2013}, (10) \citet{alencar2002}, (11) \citet{herczeg2004}, (12) \citet{ingleby2013}, (13) \citet{france14}, (14) \citet{Cleeves2015}, (15) \citet{vanBoekel2017}, (16) \citet{flaherty20}, (17) \citet{pietu2007}, (18) \citet{costigan2014}, (19) \citet{mendigutia2013}, (20) \citet{donehew2011}, (21) \citet{law21}, (22) \citet{Grady2010}, (23) \citet{rosenfeld2013b}, (24) \citet{mamajek2014}, (25) \citet{torres2006}, (26) \citet{binks2014}, (27) \citet{curran2011}, (28) \citet{avenhaus18}, (29) \citet{simon2000}, (30) \citet{andrews11}, (31) \citet{mcclure16} (32) \citet{ingleby2011}, (33) \citet{yang2012}, (34) \citet{garufi2024}, (35) This paper, (36) \citet{alcala17}, (37) \citet{mawet12}, (38) \citet{cleeves16b}, (39) \citet{seifert21} }
\end{deluxetable}

%\citet{Powell2022} find high gas diffusion, consistent with $\alpha=0.003-0.015$, around IM Lup and DM Tau and low diffusion, consistent with $\alpha\sim10^{-4}$, around HD 163296 and TW Hya. These constraints are in line with our results, with the caveat that $\alpha$ values derived from CO emission are tracing the behavior in the upper layers of the disc rather than the midplane traced by CO diffusion, where turbulence is likely lower, as discussed earlier. 

This dichotomy in turbulent velocities raises the question of what is unique about DM Tau and IM Lup, which in turn is related to the question about what could be driving the turbulence around these sources. E.g., if the turbulence is driven by the magneto-rotational instabilities then the magnetic field and ionization structure may be different between the turbulent and non-turbulent systems, while if turbulence is instead driven by the vertical shear instability, then the density and temperature structure, and the cooling timescale, are important factors. With a sample of six sources, these questions are difficult to fully answer, but we can survey our current knowledge. 

%{\it Gravito-turbulence: }
\subsection{Gravito-turbulence}
If the disc is sufficiently cold and massive then gravito-turbulence may drive turbulent gas motions. Models of gravito-turbulence predict velocities of 0.2 -- 0.4c$_s$ \citep{forgan12,shi14}, consistent with the turbulent motion around IM Lup. High resolution dust continuum images of IM Lup reveal spiral arms between 25 and 110 au, which are consistent with gravitational instabilities \citep{huang18,cadman20}, although they may alternatively be due to an embedded massive planet \citep{verrios2022}.%, while \citet{sierra21} find that it is the inner disc (r$<$50 au) that may be gravitationally unstable. 

\begin{figure}%[ht]
\center\includegraphics[scale=.42]{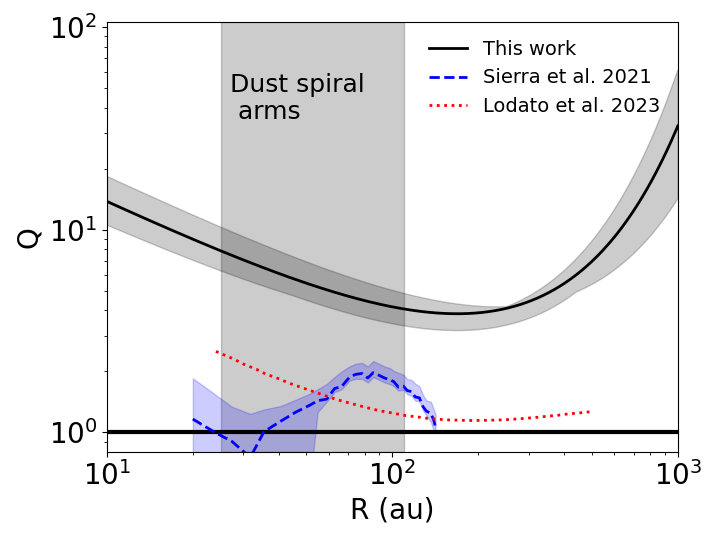}
\includegraphics[scale=.5]{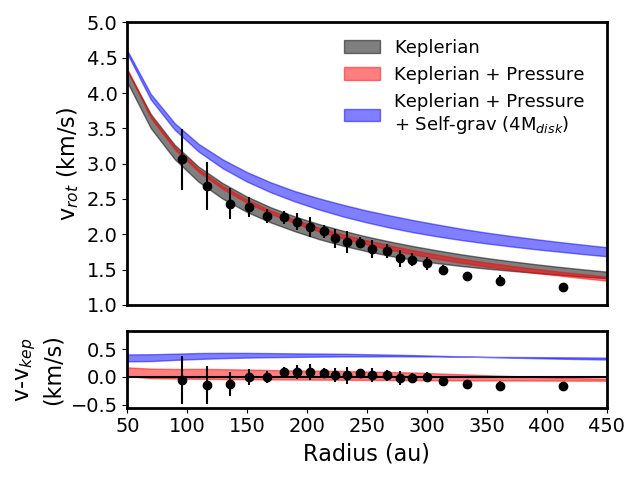}
\caption{(Top:) Toomre Q values, based on disc models derived from the CO observations. The solid curved line represents the fiducial model, while the gray band around this line demonstrates the range of Q values among the different model fits, illustrating the uncertainty in the Q estimate. The radii over which the spiral arms are seen in the dust emission are indicated by the gray shaded region. The horizontal solid line indicates the instability boundary of Q=1. Our derived disc structure lies comfortably above this boundary, indicating that the disc is likely gravitationally stable. \citet{sierra21} and \citet{Lodato2023} find lower Q values, near instability, with both results having a higher concentration of mass towards the center of the disk.
(Bottom:) Orbital velocity in CO emitting region, based on Keplerian motion (grey band, accounting for the varying CO emission heights) and Keplerian motion plus the gas pressure gradient (red band). The blue band includes Keplerian motion, the pressure gradient, and self gravity, with a disc that is massive enough that Q$\sim$1. If the disc were massive enough to be gravitationally unstable, its orbital velocity would show a detectable deviation from Keplerian motion, which is not observed \citep[][black dots]{pin18}. The drop in the observed velocities below the predictions at $R>$300 au is possibly a sign of a photoevaporative wind \citep{haworth17}. \label{figure:toomreQ}}
\end{figure}

The Toomre Q (=$c_s \Omega / \pi G\Sigma$) parameter can be used to estimate the susceptibility of the disc to gravitational collapse/instability \citep{kratter16}. With sufficiently short cooling timescales, the disk will fragment, otherwise gravito-turbulence is more likely. Given that disk fragments have not been observed around IM Lup, we assume the cooling timescale is sufficiently long that the disk would not fragment if Q$\sim$1. Figure~\ref{figure:toomreQ} shows the Toomre Q value based on the derived disc parameters from the different models. The disc reaches a minimum of Q$\sim$3, which is above the boundary where instabilities begin to set in \citep[$Q\lesssim1.5$,][]{shi14}. This is consistent with \citet{cleeves16b} who found that Q reaches a minimum of $\sim$4 at 70 au. Assuming the temperature profile and surface density shape derived from CO, the disc mass would need to be $\gtrsim$0.7M$_{\odot}$ in order for the disc to be gravitationally unstable. 

Our ability to rule out gravito-turbulence depends on the assumed radial profile of the disc mass, and the temperature profile. As noted earlier, the midplane temperature may be smaller than derived in our fiducial model, but since the Toomre Q depends on $\sqrt{T}$ the midplane temperature would need to be an order of magnitude smaller for the disc to be gravitationally unstable. \citet{sierra21} find that the radii $<$20 au are gravitationally unstable, based on a dust mass derived from multi-frequency analysis of radio continuum images, and assuming a gas-to-dust ratio of 100. \citet{Lodato2023} use the CO velocity profile to derive Q$\sim$1-2 throughout much of the disc, a factor of a few lower than our result despite similar disc masses, due to a much smaller value of $R_c$ (66 vs 248 au here) and a much steeper temperature profile ($q$ = -0.66 vs. -0.355 here). This raises the density in the inner disc, and decreases the temperature in the outer disc, both of which contribute to lowering the Toomre Q throughout the disc. While we have narrowed the region of parameter space (long cooling timescale, to prevent fragmentation, compact disk, steep temperature profile) needed for gravitational instabilities to be present, more direct tracers of the surface density and midplane temperature profile are needed to confidently determine whether or the disc around IM Lup is gravitationally unstable.

In general, disc mass (Table~\ref{table:stellar_params}) does not stand out as a substantial difference between the turbulent sources (IM Lup: M$_{\rm gas}$=0.175 M$_{\odot}$, DM Tau: M$_{\rm gas}$=0.04 M$_{\odot}$) and the non-turbulent sources (MWC 480: M$_{\rm gas}$ = 0.046 M$_{\odot}$, V4046: M$_{\rm gas}$ = 0.09 M$_{\odot}$, TW Hya: M$_{\rm gas}$ = 0.05 M$_{\odot}$, HD 163296: M$_{\rm gas}$ = 0.09 M$_{\odot}$). This suggests that gravito-turbulence is not a substantial source of turbulence at these ages even in many of the brightest systems. 

%This result depends on the assumed disc mass and while disc masses are difficult to directly estimate from e.g. dust continuum emission \citep{Andrews20}, a massive disc will leave an imprint on the orbital velocities within the disc, which may be detectable \citep{veronesi21}.
This result depends on the assumed disc mass, as well as the surface density and temperature profiles through the disc. While disc masses are difficult to directly estimate from e.g. dust continuum emission \citep{Andrews20}, a massive disc will leave an imprint on the orbital velocities within the disc, which may be detectable \citep{verrios2022}.
The lower panel of Figure~\ref{figure:toomreQ} shows the velocity profile, based on our fiducial model, when including the velocity corrections for the pressure gradient and self-gravity for a gravitationally unstable disc (M$_{\rm gas}$ = 0.7 M$_{\odot}$). The effect of self-gravity is calculated using the method in \citet{veronesi21}, based on the derivations by \citet{bertin99} and \citet{lodato07}. Under self-gravity the orbital frequency becomes:
\begin{equation}
    \Omega^2 = \frac{GM_*}{(R^2+z^2)^{3/2}} + \frac{1}{R}\frac{1}{\rho}\frac{dP}{dR} + \frac{1}{R}\frac{d\Phi_{\sigma}}{dR}(R,z)
\end{equation}
where the last term represents the contribution from disk self-gravity. Here $\Phi_{\sigma}$ is the disk contribution to the gravitational potential, which is calculated via:
\begin{equation}
\begin{split}
    \frac{\partial \Phi_{\sigma}}{\partial R}(R,z) = \frac{G}{R}\int^{\infty}_{0} [K(k) - \frac{1}{4}\left(\frac{k^2}{1-k^2}\right) \\
    \times\left(\frac{R'}{R}-\frac{R}{R'}+\frac{z^2}{RR'}\right )E(k)] \sqrt{\frac{R'}{R}}k\Sigma(R')dR'
    \end{split}
\end{equation}
where $k^2=4RR'/[(R=R')^2+z^2$ and $E(k)$ and $K(k)$ are complete elliptic integrals of the first find. If the disc were massive enough to be gravitationally unstable, it would exhibit orbital velocities $\sim$0.4 km s$^{-1}$ greater than the Keplerian motion. \citet{pin18} measure sub-Keplerian velocities, deviating from Keplerian motion by $\sim$0.15 km s$^{-1}$, that are inconsistent with the profile expected for a massive disc. The observed velocity profile could be matched with a smaller stellar mass, although this would require M$_{*}\sim$0.6M$_{\odot}$ with M$_{\rm disc}\sim$0.7 M$_{\odot}$, in contrast to M$_{*}$=1.1M$_{\odot}$ and M$_{\rm disc}=$0.175 M$_{\odot}$ used in our models. We can confidently rule out such a disc to star mass ratio because in that case the signatures of gravitational instability would not be subtle \citep{kratter16}, and such a low stellar mass has been ruled out by modeling of the disc kinematics \citep{teague21,Lodato2023,Izquierdo2023}. This indicates that gravito-turbulence is unlikely to be the driver of turbulence around IM Lup. Accounting for disc self-gravity in our modeling of the CO emission, following the velocity prescription of \citet{veronesi21}, we find that disc self-gravity does not substantially bias our estimate of the turbulence (Table~\ref{table:results}).

%{\it Magneto-Rotational Instability: }
\subsection{Magneto-Rotational Instability}
In the context of the magneto-rotational instability, the key factors are sufficient ionization and sufficiently strong vertical magnetic fields \citep{simon18}. The exact conditions under which MRI is active \citep[e.g.][]{bai2011,Flock2012} depend on the particulars of the system (e.g., density and temperature) as well as the region of the disk, and the associated non-ideal MHD effects operating (e.g. Ohmic dissipation, Hall effect, Ambipolar diffusion). As an example of the conditions under which MRI could operate, \citet{simon18} find that for magnetic field strengths greater than 5-10 $\mu$G, in the presence of ionization from FUV photons, the MRI would produce detectable ($>$0.1c$_s$) turbulence in the ambipolar dominated outer disk. The ionization comes from a combination of FUV, X-ray, and cosmic ray flux with models suggesting that FUV flux plays a larger role in the CO emitting layer \citep{perez-becker11,simon18}. Typical FUV luminosities of young stellar objects range from 2$\times$10$^{-6} - 0.2$ L$_{\odot}$ \citep{france14}. There is no clear trend in FUV luminosity among our sample (Table~\ref{table:stellar_params}), with the non-turbulent HD 163296 having the highest FUV luminosity \citep{meeus12}, followed by turbulent IM Lup \citep{cleeves16b}, non turbulent TW Hya and V4046 Sgr \citep{france14}, and then turbulent DM Tau having the weakest FUV luminosity \citep{yang12,france14}. 

One caveat in this comparison is that the FUV flux comes from accretion shocks \citep{france14}, which are highly variable (e.g., the accretion rate onto HD 163296 increased by a factor of 10 in the 15 years between observations \citet{mendigutia2013}). This, in turn, may lead to time-variable disc ionization \citep{cleeves17}. The FUV observations are snapshots in time, which may not reflect the ionizing emission at the time when the turbulence was measured.

%Ardila et al. 2013
%Yang et al. 2012
%France et al. 2014b

\begin{figure*}
\center
\includegraphics[scale=.5]{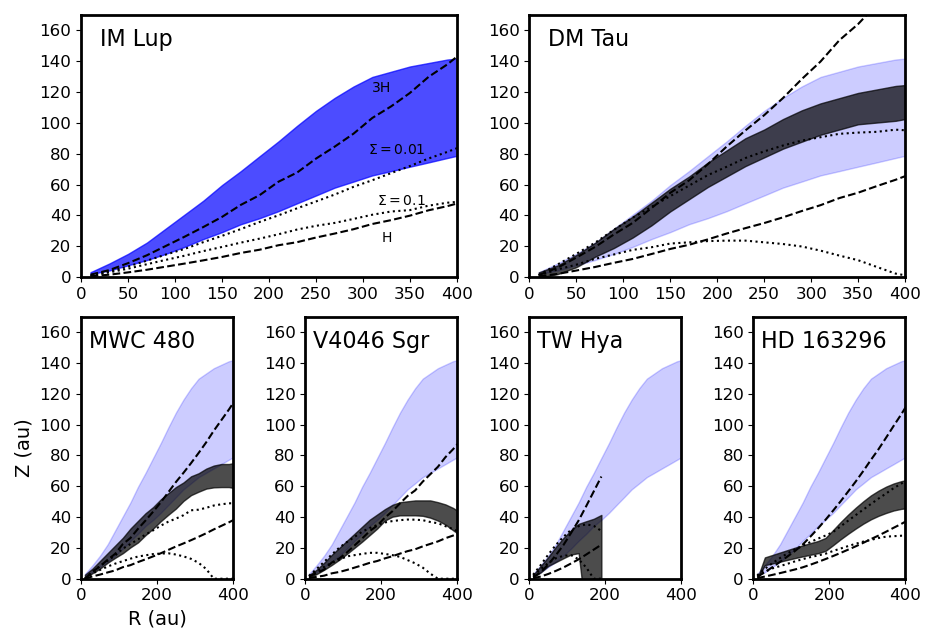}
\caption{Grey bands represent the heights between which 50\% of the flux for CO(2--1) arises at a given radius (as described in Section~\ref{sec:results}), for the different systems (the blue band in each panel is the CO(2--1) emitting region around IM Lup). The dashed lines indicate the pressure scale height, $H$, and three times the pressure scale height. The dotted lines indicate surface density boundaries of $\Sigma=0.01$ g cm$^{-2}$ and 0.1 g cm$^{-1}$. While the emitting regions around DM Tau and IM Lup are systematically higher relative to the sources with weak turbulence, they are not systematically higher relative to the pressure scale height, or the column density within the disc.\label{figure:emission_heights}}
\end{figure*}

Another caveat is that these observations measure the FUV luminosity at the stellar surface, while the FUV flux that reaches the outer disc is more important for influencing turbulence. High-energy photons traveling outwards could be blocked by an inner disc wind \citep{pascucci20}, preventing a high stellar FUV flux from ionizing the outer disc. Similarly, a stellar wind can exclude cosmic rays, decreasing the ionization fraction in the disc \citep{Cleeves2015}. Assuming they are unimpeded in their journey to the outer disc, FUV photons penetrate down to $\Sigma=0.01-0.1$ g cm$^{-2}$, depending on the FUV luminosity, grain abundance, and gas composition, while X-ray photons and cosmic rays can ionize gas closer to the midplane \citep{perez-becker11}. The CO emitting heights for our sample are between 0.01 g cm$^{-2}$ and 0.1 g cm$^{-1}$ (Figure~\ref{figure:emission_heights}), which, assuming similar levels of irradiating flux, will result in similar ionization structures. As mentioned above, these systems exhibit a range of FUV luminosities and if the depth of ionized region follows the FUV luminosity we would expect ionization to extend deeper into the disks around HD 163296 and IM Lup, and be confined to the largest heights around DM Tau.

Ionized molecules, such as HCO$^+$, N$_2$H$^+$ and their isotopologues, have been detected around all of our sources \citep{teague15,cataldi21,aikawa21} indicating that at some level the outer discs around these targets contain ionized gas. Whether the disc is sufficiently ionized for MRI requires more detailed analysis of multiple molecular species. \citet{seifert21} find high cosmic ray ionization ($\gtrsim10^{-17}$ s$^{-1}$) beyond 80 -- 100 au in the disc around IM Lup such that the region beyond 100 au is feasibly MRI active, consistent with our detection of non-zero turbulence. \citet{aikawa21} find that the ionization rate is higher around IM Lup than around HD 163296 and MWC 480, consistent with the relative turbulence levels between these targets. \citet{Cleeves2015} find a cosmic ray ionization rate $\lesssim10^{-19}$ s$^{-1}$ around TW Hya, again consistent with the weak turbulence around this target (Table~\ref{table:stellar_params}). %On the other hand, \citet{trapman22} find a high cosmic ray ionization rate around TW Hya ($5\times10^{-18}$ -- $10^{-17}$ s$^{-1}$) and a weak cosmic ray ionization rate around DM Tau ($<10^{-18}$ s$^{-1}$), contrary to the turbulent velocities... 

%\citet{trapman22} find that the N$_2$H$^+$/C$^{18}$O ratio around DM Tau is consistent with midplane cosmic ray ionization rates of 10$^{-18}$ -- 10$^{-19}$ s$^{-1}$, similar to that around HD 163296 and IM Lup \citep{aikawa21}. On the other hand, \citet{Cleeves2015} find a cosmic ray ionization rate $\lesssim10^{-19}$ s$^{-1}$ around TW Hya. 

%Fuji & Kimura 2022: Ionization rate in the outer disc of IM Lup, beyond 100 au, increases by an order of magnitude.
%[from Jake] One comment that I couldn’t fit onto the page is regarding the first full paragraph on page 11.  We first say that Seifert+21 finds a CR ionization rate of 1.e-17 (or larger), but then you say that DM tau has a smaller CR rate (1.e-19-1.e-18 /s <— by the way, these are listed in descending order in the paper, which is not the convention). You say this is similar to that around HD 163296, which makes sense since that one had weaker turbulence, but then you say that this is also similar to that in IM Lup.  But it’s not — IM Lup is about an order of mag. higher AND obviously has stronger turbulence.  Finally, I don’t understand how the comparison with TW Hya fits in — TW Hya had only an upper limit on the turbulence as well and has a smaller CR rate.  So, why do we say “On the other hand”.

The magnetic field strength remains unknown for planet-forming discs, and could vary enough to explain the difference in turbulence strengths. Simulations of ideal MHD \citep{hawley1995,simon2009} and ambipolar diffusion in an unstratified disk \citep{bai2011} find turbulent velocity to be correlated with magnetic field strength, but this trend is not as clear in stratified non-ideal MHD simulations \citep{simon18}. Broadly speaking, a sufficiently strong magnetic field is needed to activate the MRI, and in this way turbulence is related to magnetic field strength. Recent ALMA observations of line polarization find upper limits of a few mG on the background vertical magnetic field \citep{vlemmings19,harrison21}, well above the $\sim10\mu$G regime where MRI operates \citep{simon18}. DM Tau and IM Lup, at $\sim$1 Myr, are younger than the other systems (Table~\ref{table:stellar_params}) and the magnetic field strength may change over time if the magnetic field is diffused outwards, and/or advected inwards \citep{bai2017,simon18,cui2021}. Paleomagnetic measurements within our solar system indicate magnetic fields, of unknown orientation, of $\sim$0.5 - 1 G between 1 and 3 au and $>$ 0.06 G in the outer solar system when our solar system was 2 Myr old, with non-detections with an upper limit of $<$0.006 G and $<$0.003 G after about 3.9 Myr and 4.9 Myr for the inner/outer solar system \citep{weiss21}. This is consistent with an age dependence for turbulence, assuming all young stars start with similar magnetic field strengths. Recent studies of the vertical extent of mm dust in class 0 and I sources \citep{villenave2023,lin2023,guerra-alvarado2023} find that these young systems have thicker disks than class II disks, consistent with stronger turbulence at younger ages (but see \citet{lim2023} for a discussion of how smaller dust scale heights may not necessarily equate to weaker turbulence). 

Age is certainly not completely deterministic in setting turbulence. HL Tau is, as a class I object, much younger than DM Tau and IM Lup, but shows weak turbulence \citep{pinte2016}, as does the class I/II source DG Tau \citep{ohashi2023}. Similarly, many class II discs with ages similar to DM Tau and IM Lup have gas disc radii consistent with $\alpha=10^{-3}-10^{-4}$ \citep{najita18,trapman20,long2022}. Source-to-source variations in the initial magnetic field strength could result in some systems being MRI active while others are not; observations of magnetic field in protostars are still limited, but they find a variety of strengths \citep{Tsukamoto2023} and orientations \citep{Hull2014,Yen2021}. More work is needed to understand the strength of magnetic fields in planet-forming discs, and how it evolves with time. 

%Ohashi et al. 2023 find dust settlinng around DG Tau. Is  DG Tau class I or class  II??

%{\it Hydrodynamic Instabilities: }
\subsection{Hydrodynamic Instabilities}
If the magnetic field or ionization is sufficiently weak then hydrodynamic instabilities may play a larger role. The Vertical Shear Instability (VSI) is a hydrodynamic instability that is related to the change in orbital motion with height within the disc ($\Omega\sim(r^2+z^2)^{-3/2}$), and is  expected to play a larger role relative to other hydrodynamic instabilities at the large radii that ALMA observations are sensitive to \citep{lyra2021,Lesur2022}. Velocities driven by VSI, which are dominated by vertical motions \citep[e.g.,][]{flores-rivers2020}, can reach up to $\sim$100 m s$^{-1}$ in the upper layers of the outer disc \citep{flock17,barraza-alfaro2021}, increasing by a factor of 10 between the midplane and $z/r\sim$0.3 \citep{flock17,pfeil2021}. 

%The CO emitting regions around the turbulent systems DM Tau and IM Lup occur at $z/r\sim0.2-0.4$, as opposed to $z/r\sim0.1$ for the non-turbulent systems \citep[Figure~\ref{figure:emission_heights}][]{pinte08,law21,Law2022}, consistent with VSI\footnote{MRI also predicts varying turbulence with height within the disc, assuming sufficient ionization and magnetic field strength \citep[e.g.][]{simon15}}.  

%z/r~0.3-0.4 for DM Tau, ~0.25 for IM Lup
%z/r~0.1-0.2 for MWC 480, ~0.1 for V4046, ~0.1-0.2 for TW Hya, ~0.1 for HD 163296

\begin{figure*}
\centering
\includegraphics[scale=.7]{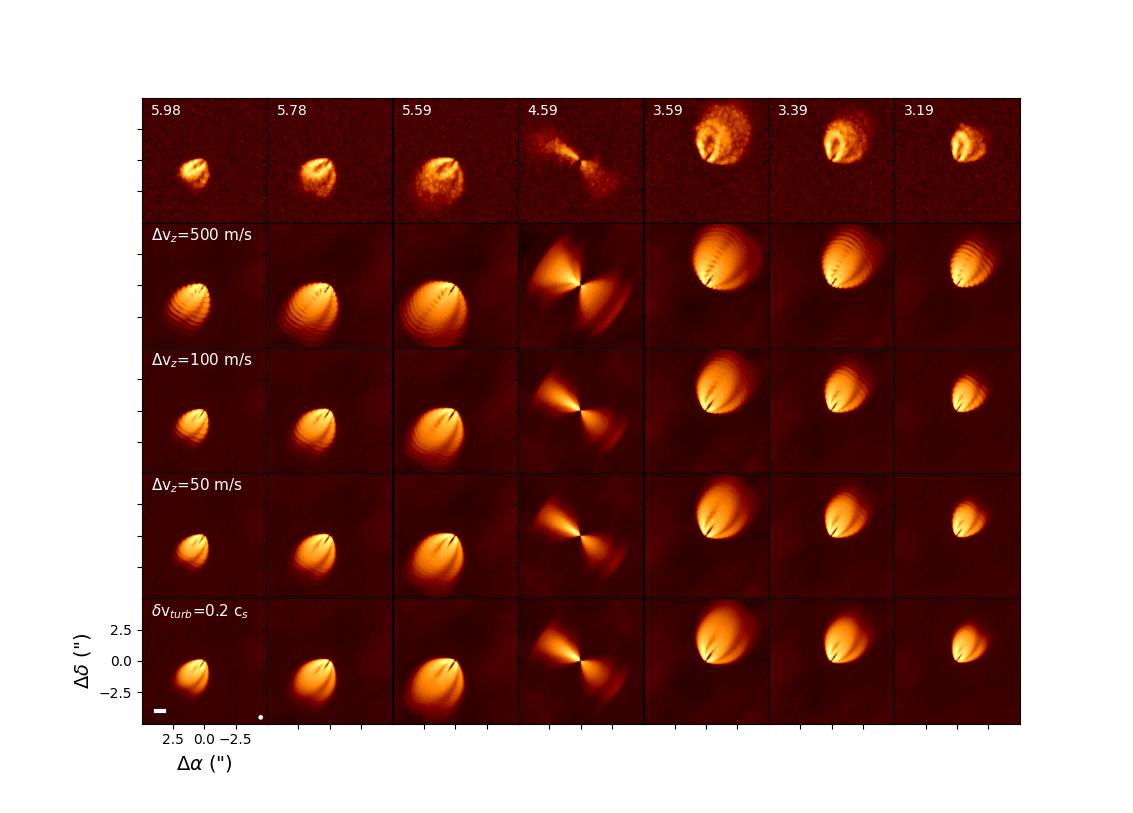}
\caption{Observed CO(2--1) channel maps (top row, \citet{oberg21}) as compared to simple models of VSI-like motion (2nd, 3rd, and 4th rows, $\Delta v_z\sin(R/(60 {\rm au}))$), and isotropic turbulence (bottom row). The VSI-like models produce corrugated features that are visible in the outer disc. These become increasing difficult to detect as the velocity scale approaches that of the observed turbulence around IM Lup ($\delta v\sim50-100$m s$^{-1}$). There is no strong evidence for VSI-like behavior in the disc around IM Lup, although it may be below the detection limit of the data.\label{figure:maps_vsi}}
\end{figure*}

VSI produces velocities whose amplitude oscillates with a characteristic length scale of $\sim$0.1R \citep{flock20,barraza-alfaro2021}. At small radii, and with low-resolution data, these features are unresolved and would contribute to line broadening in a way that is indistinguishable from  isotropic random non-thermal motion, but at larger radii these structures are potentially observable \citep{barraza-alfaro2021} with high-resolution ALMA observations. For the data used here, with a resolution of $\sim$60 au, radii less than $\sim$600 au would likely encompass VSI features moving in opposite directions, while higher resolution data from MAPS \citep{oberg21}, with a spatial resolution of $\sim$10au, would only smear together VSI features at radii smaller than $\sim$100 au. In figure~\ref{figure:maps_vsi} we show the MAPS CO channel maps \citep{oberg21}, along with simple models in which a vertical velocity component has been added, that varies as a sinusoidal function of radius with a wavelength of 60 au (=$\delta v_z\sin(R/(60 {\rm au}))$). At a spatial resolution of $\sim$10 au, the MAPS data can spatially resolve VSI-like features in the outer disc and our simple VSI models produce corrugated structures in the channel maps, similar to those seen in more detailed models \citep{barraza-alfaro2021}. Corrugated features are not strongly evident in the MAPS data, but they are difficult to observe at the small velocities ($\sim$50-100 m s$^{-1}$) expected for IM Lup. No corrugated features are evident in the residuals either (Figure~\ref{figure:imdiff}). 

The highly anisotropic nature of VSI ($v_z\gg v_r$) is very efficient at lofting dust grains in to the disc atmosphere \citep{flock17}, consistent with the large dust scale height derived from the IM Lup scattered light image, but may be challenged by the small scale height seen in the millimeter continuum \citep{Franceschi2022}. While larger dust grains are expected to more effectively settle towards the midplane, the VSI may still be effective at lofting up these grains \citep{dullemond2022} The dust scale height is directly dependent on the vertical component of the motion, while our observations of the gas motion, under our assumption of isotropic turbulence, would only be sensitive to the line of sight component of this vertical motion.

VSI as the source of the turbulence is challenged by the large disc radius around IM Lup (Table~\ref{table:stellar_params}), which is consistent with a viscous stress of $\alpha\sim10^{-2}$ \citep{trapman20,long2022}, much larger than produced by the VSI \citep[$\alpha_{\rm visc}=1.5\times10^{-4}$,][]{flock20}. VSI may not exist in isolation, and may operate in concert with MRI and a wind \citep{Cui2022}. MRI, as well as a wind \citep{Yang2021}, may contribute to the radial spreading of the disc in the presence of VSI, leading to the observed large disc radii for IM Lup and DM Tau.

In the context of the VSI, turbulence is generated when the cooling timescale is much shorter than the orbital period \citep{Lesur2022}. The cooling timescale in turn depends on the dust population \citep{umurhan2017,lyra2021}, since the large opacity of the dust relative to the gas makes it efficient at cooling the disc. If e.g., dust were present at large radii around DM Tau and IM Lup and not the other sources then this would, at least qualitatively, indicate a shorter cooling timescale around DM Tau and IM Lup. Scattered and polarized light images, sensitive to the small dust grains present in the outer disc, find emission on size scales (Table~\ref{table:stellar_params}) comparable to the CO emission for all of our sources \citep{Wisniewski2008,Grady2010,vanBoekel2017,avenhaus18,garufi2024}. This indicates that there is no obvious difference in the radial dust distribution between the turbulent and non-turbulent discs, although detailed work is needed to calculate the exact cooling timescales in the outer disc, in order to confirm whether or not the conditions are sufficient for the VSI. 

\section{Conclusions}
We find significant turbulence around IM Lup, at a level of (0.18 -- 0.30)c$_s$, based on CO(2--1), $^{13}$CO(2--1), and C$^{18}$O(2--1) observations. This result is robust against various systematic effects (e.g., flux calibration, midplane temperature uncertainties, disc self-gravity) and is consistent with the unusually large CO disc radius around IM Lup \citep{trapman20,long2022} and the non-thermal broadening of CN \citep{paneque-carreno2023}. The turbulence level is similar to that around DM Tau ((0.25 -- 0.33)c$_s$), but stronger than those around MWC 480, V4046 Sgr, TW Hya, and HD 163296 \citep{Teague18,flaherty20}. %A comparison between the turbulence derived from CO and measures of dust settling are consistent with a decrease in turbulence from the surface layers down to the midplane of the disc. 

The exact reason for the different turbulence levels among these sources, and the physical origin of the turbulence, is not entirely clear, although there are some hints in this small sample. We can rule out gravito-turbulence in the case of IM Lup, assuming the density and temperature structure derived from the CO observations, as it would impart a detectable signal on the orbital velocities of the CO gas around IM Lup. The disc around IM Lup is sufficiently ionized to be susceptible to the MRI \citep{seifert21}, although the unknown magnetic field strength makes it difficult to assess whether or not the MRI is active in this system. The young ages of the turbulent DM Tau and IM Lup discs are consistent with a scenario in which the magnetic field strength evolves with time \citep{simon18}, similar to that seen in our solar system \citep{weiss21}, although more work is needed to confirm this conclusion. High resolution, deep observations could reveal disc structures driven by the VSI \citep{barraza-alfaro2021}, or spatially varying turbulence \cite[e.g.,][]{Bosman2023}. The high-resolution MAPS observations have the potential to reveal such structures, as well as to explore e.g. radial variations in turbulence. Such an analysis requires a detailed accounting for variations in the density, temperature, and CO abundance structure beyond the parametric forms employed here. While further work is needed to understand the mechanism driving turbulence, and the demographics of turbulence among a broader sample, the our measurement of turbulence around IM Lup provides valuable clues as to the conditions under which planets can form. 

%\begin{itemize}
%    \item We find significant turbulence around IM Lup, at a level of (0.18 -- 0.30)c$_s$, based on CO(2--1), $^{13}$CO(2--1), and C$^{18}$O(2--1) observations. {\bf We are able to tightly constrain turbulence despite excluding regions of the spectrum contaminated by cloud absorption.}
%    \item This result is robust against various systematic effects (e.g., flux calibration, midplane temperature uncertainties, disc self-gravity) and is consistent with the unusually large CO disc radius around IM Lup \citep{trapman20,long2022}. The turbulence level is similar to that around DM Tau ((0.25 -- 0.33)c$_s$), but stronger than those around MWC 480, V4046 Sgr, TW Hya, and HD 163296 \citep{Teague18,flaherty20}.
%    \item 
%\end{itemize}

%although this remains challenging given the small velocity scales involved ($\sim$50-100 m s$^{-1}$).

%The large CO emitting heights around DM Tau and IM Lup, as compared to the non-turbulent sources, are consistent with VSI-driven turbulence \citep{flock20}, but whether or not the cooling timescale is short enough for the VSI to operate in these systems is still unknown.  

\section*{Acknowledgements}

We thank the referee for a detailed reading of the paper that improved the analysis and presentation. AMH is supported by a Cottrell Scholar Award from the Research Corporation for Science Advancement. We thank the referee for detailed comments that improved the paper. 

This paper made use of the following ALMA data: ADS/JAO.ALMA\#2013.1.00798.S. ALMA is a partnership of ESO (representing its member states), NSF (USA), and NINS (Japan), together with NRC (Canada), {\it MOST} and ASIAA (Taiwan), and KASI (Republic of Korea), in cooperation with the Republic of Chile. The Joint ALMA Observatory is operated by ESO, AUI/NRAO, and NAOJ. The National Radio Astronomy Observatory is a facility of the National Science Foundation operated under cooperative agreement by Associated Universities, Inc. This project has received funding from the European Research Council (ERC) under the European Union's Horizon 2020 research and innovation programme under grant agreement No 716155 (SACCRED). This work was also supported by the NKIFIH excellence frant TKP2021-NKTA-64.

This work has made use of data from the European Space Agency (ESA) mission {\it Gaia} (\url{https://www.cosmos.esa.int/gaia}), processed by the {\it Gaia} Data Processing and Analysis Consortium (DPAC, \url{https://www.cosmos.esa.int/web/gaia/dpac/consortium}). Funding for the DPAC has been provided by national institutions, in particular the institutions participating in the {\it Gaia} Multilateral Agreement.

%\facilities{ALMA}

%\software{astropy \citep{astropy13},  
%          RADMC-3D \citep{dullemond12},
%          galario \citep{Tazzari2018},
%          emcee \citep{foreman-mackey13},
%          CASA \citep{mcmullin2007}}

%%%%%%%%%%%%%%%%%%%%%%%%%%%%%%%%%%%%%%%%%%%%%%%%%%
\section*{Data Availability}

The ALMA data used in this paper are from project code 2013.1.00798.S and are available in the ALMA archive. The derived data (e.g., cleaned maps) as well as the models, and code used to generate them, will be shared on reasonable request to the corresponding author. 
 
%The inclusion of a Data Availability Statement is a requirement for articles published in MNRAS. Data Availability Statements provide a standardised format for readers to understand the availability of data underlying the research results described in the article. The statement may refer to original data generated in the course of the study or to third-party data analysed in the article. The statement should describe and provide means of access, where possible, by linking to the data or providing the required accession numbers for the relevant databases or DOIs.



\bibliographystyle{mnras}
\bibliography{example} % if your bibtex file is called example.bib

%\begin{thebibliography}{99}
%\end{thebibliography}

%%%%%%%%%%%%%%%%%%%%%%%%%%%%%%%%%%%%%%%%%%%%%%%%%%

%%%%%%%%%%%%%%%%% APPENDICES %%%%%%%%%%%%%%%%%%%%%

\appendix

\section{Channel Maps}
Full channel maps, showing the data, model, and residuals, for the fiducial model. Channels with cloud contamination have been excluded.

\begin{figure*}
\centering
\includegraphics[scale=.31]{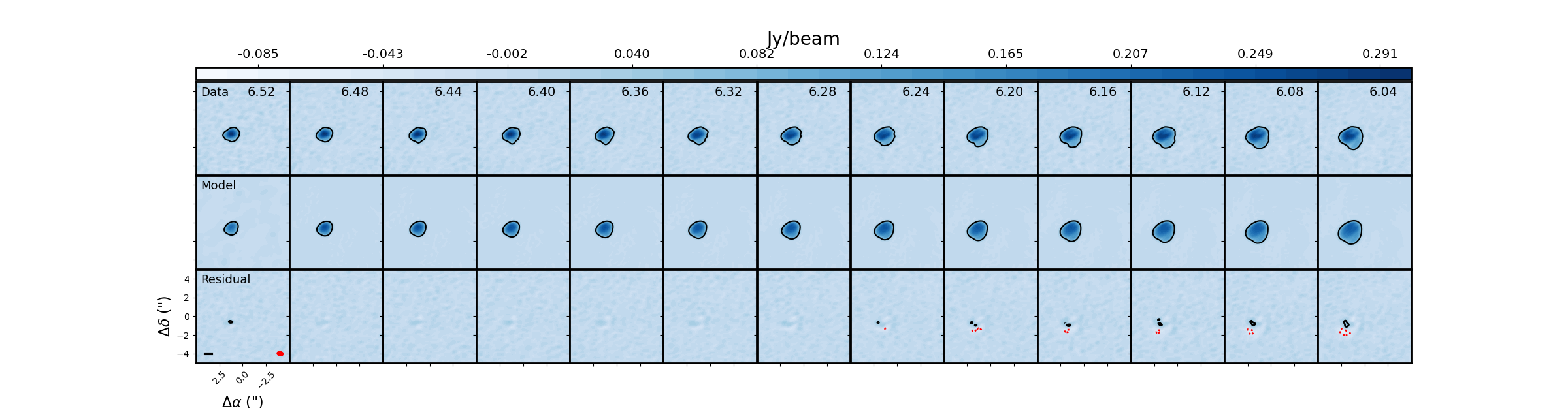}
\includegraphics[scale=.31]{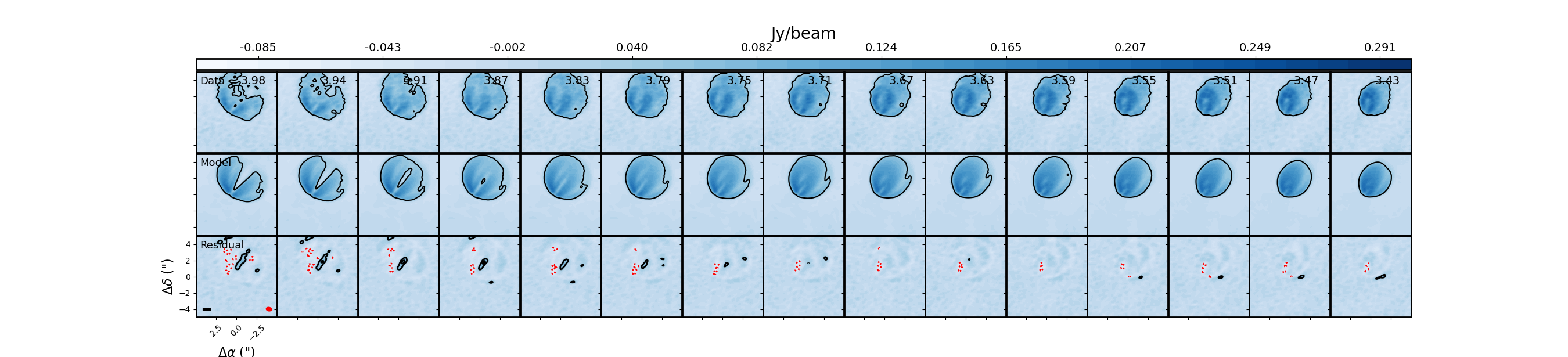}
\includegraphics[scale=.31]{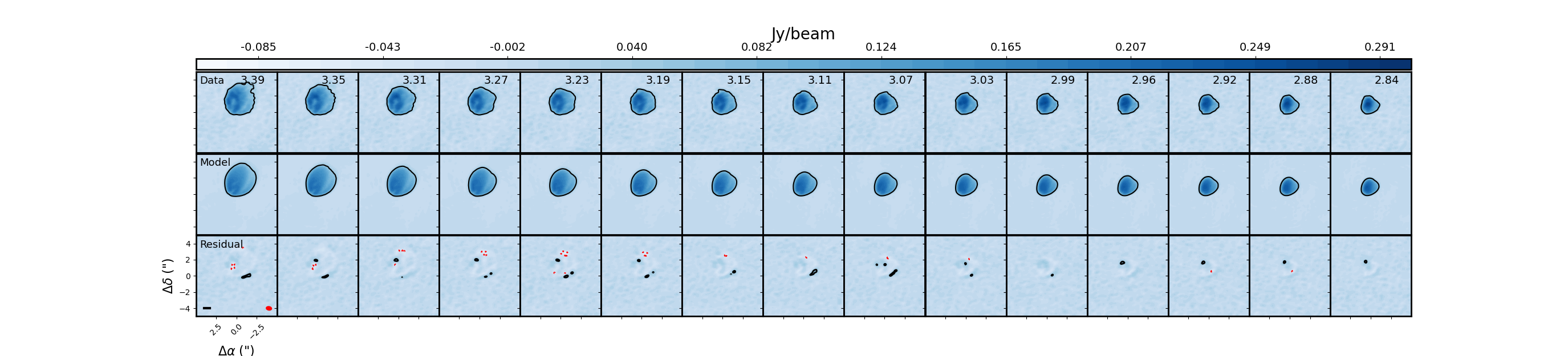}
\includegraphics[scale=.31]{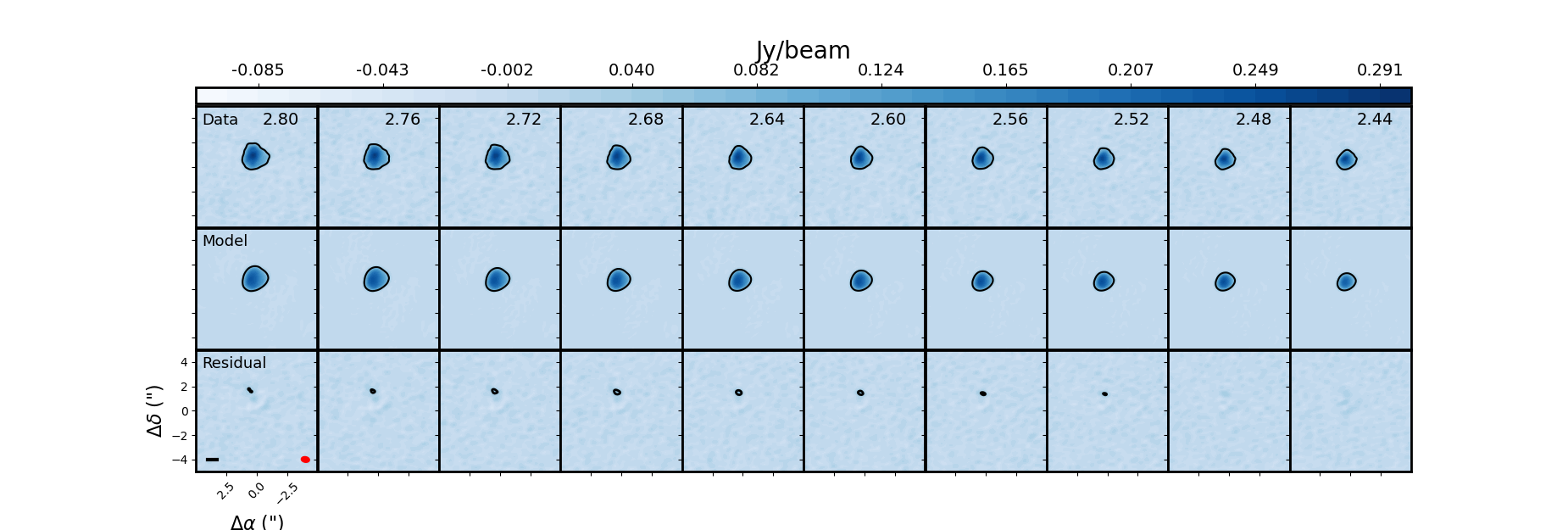}
\caption{Channel maps of the data, the fiducial model, and the residuals. In the data and model channel maps the contour indicates the 5$\sigma$ level ($\sigma=9$ mJy/beam). In the residual maps, positive (black) and negative (red dashed) contours are at levels in multiples of 5$\sigma$. Lower left panels include a 100 au scale bar and the beam shape. Channels between 4 and 6 km s$^{-1}$ have been removed due to cloud contamination.} 
\end{figure*}

%%%%%%%%%%%%%%%%%%%%%%%%%%%%%%%%%%%%%%%%%%%%%%%%%%
\section{Cloud Contaminated Channel Maps}
Channel maps for the channels that we have excluded because of cloud absorption, showing the data, model, and residuals, for the fiducial model.
\begin{figure*}
\centering
\includegraphics[scale=.31]{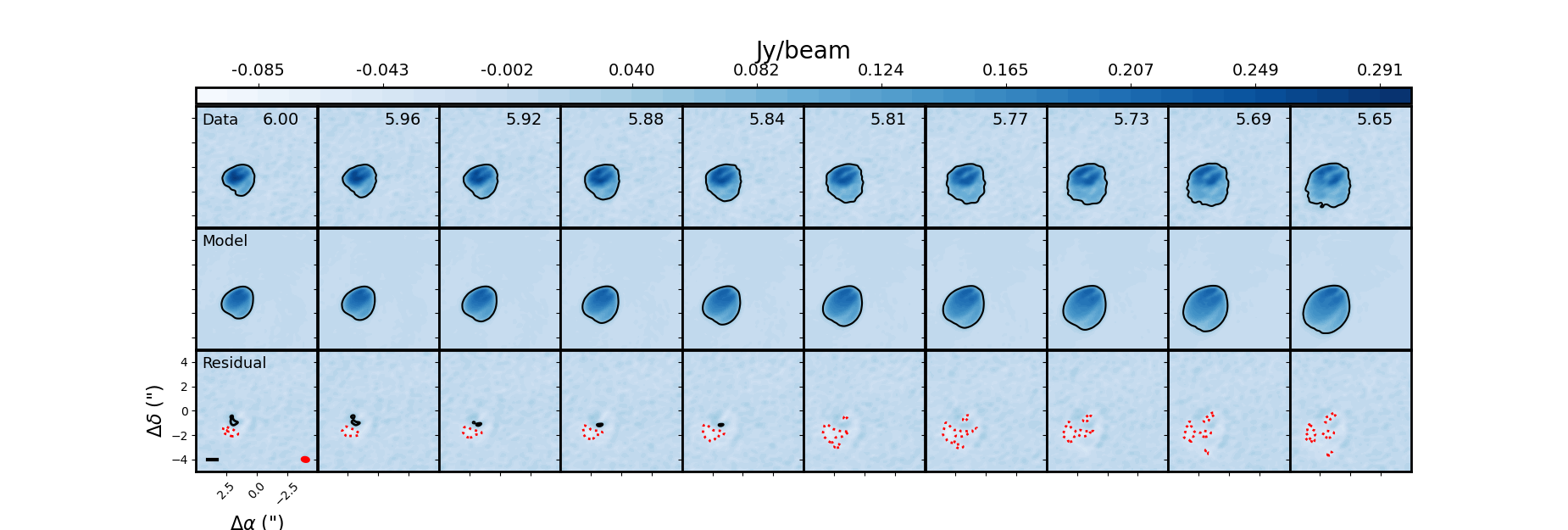}
\includegraphics[scale=.31]{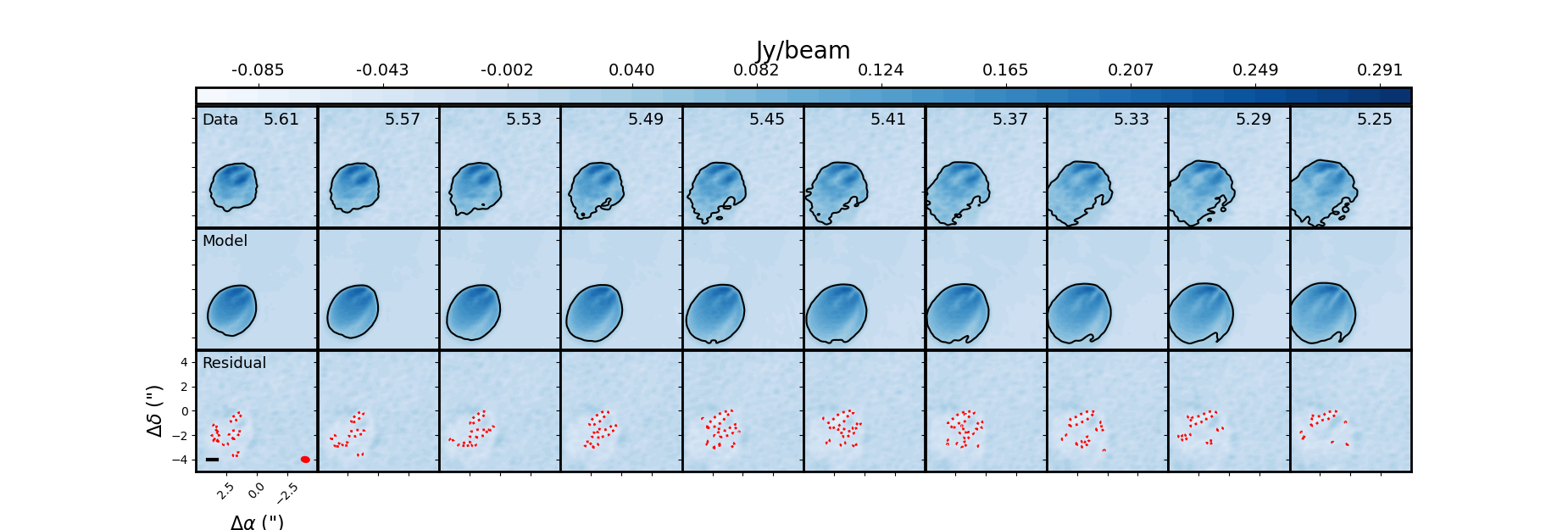}
\includegraphics[scale=.31]{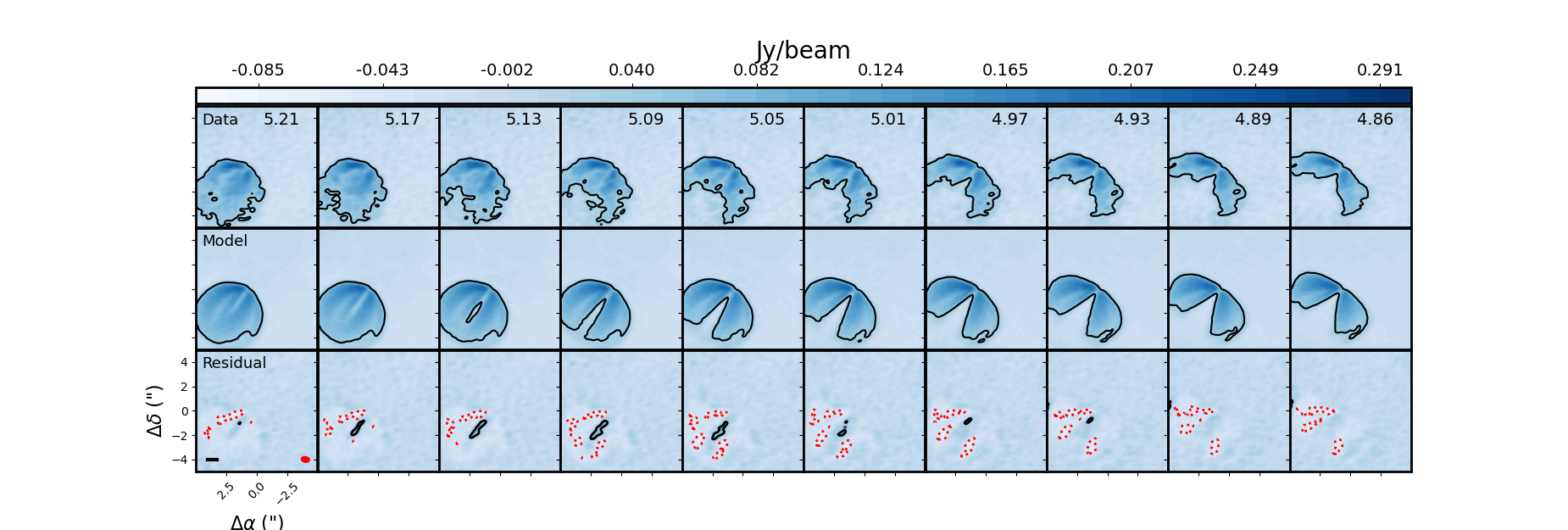}
\includegraphics[scale=.31]{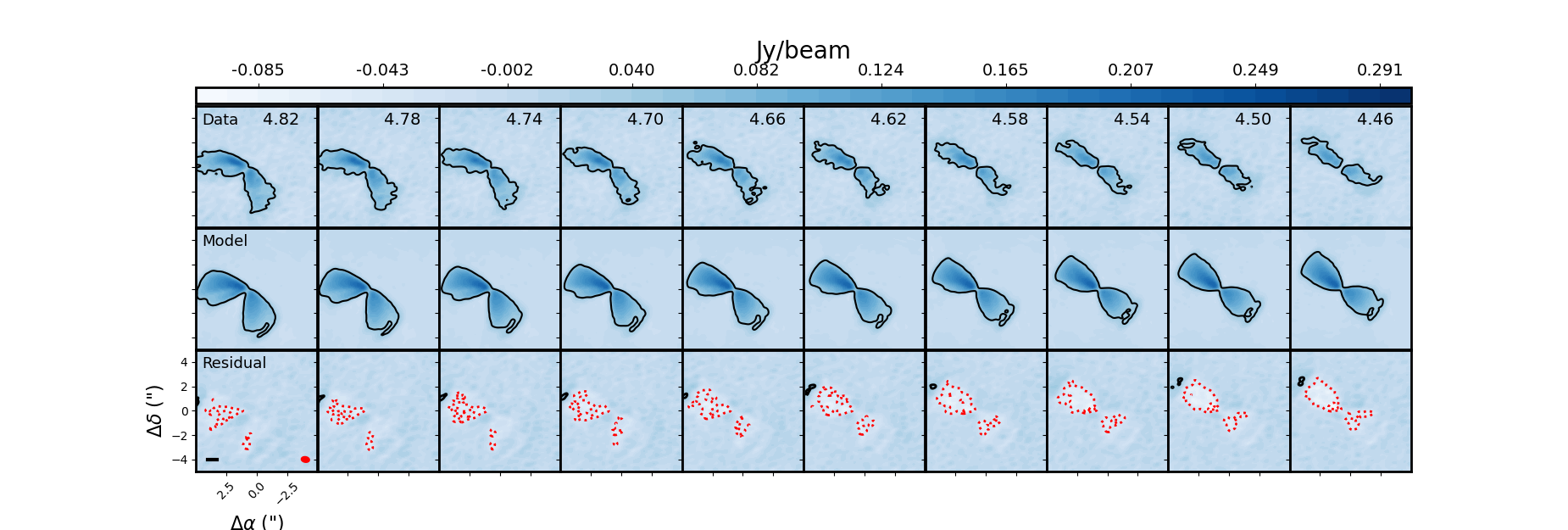}
\includegraphics[scale=.31]{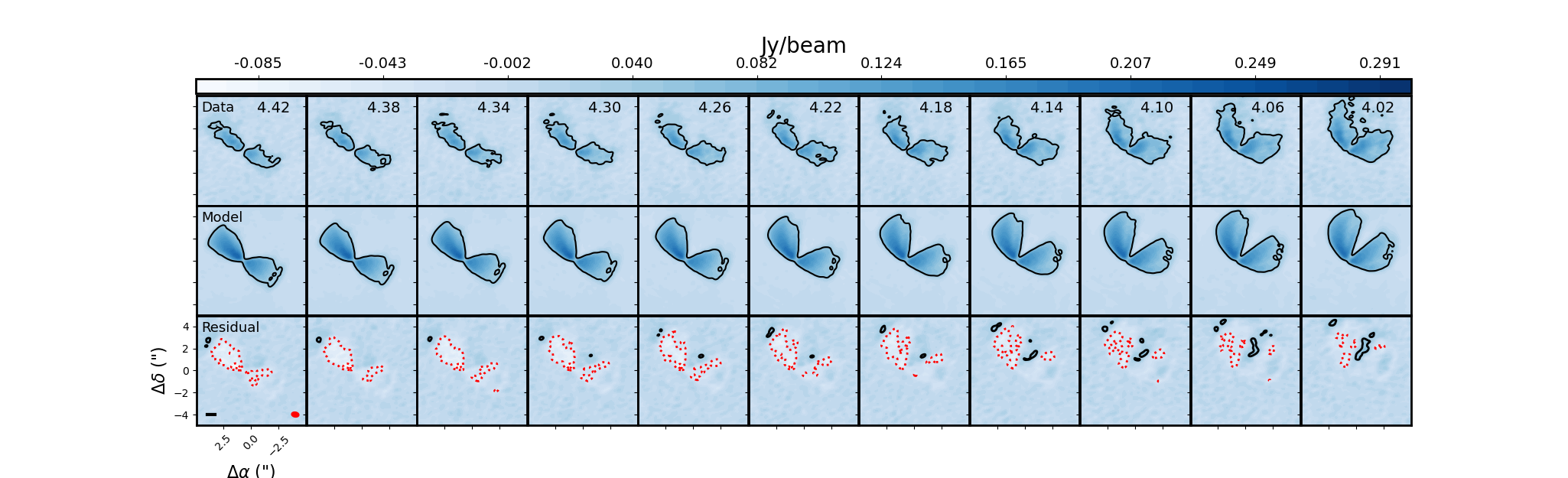}
\caption{Channel maps of the data, the fiducial model, and the residuals. In the data and model channel maps the contour indicates the 5$\sigma$ level ($\sigma=9$ mJy/beam). In the residual maps, positive (black) and negative (red dashed) contours are at levels in multiples of 5$\sigma$. Lower left panels include a 100 au scale bar and the beam shape. Channels between 4 and 6 km s$^{-1}$ have been removed due to cloud contamination.} 
\end{figure*}

% Don't change these lines
\bsp	% typesetting comment
\label{lastpage}
\end{document}